\documentclass{aa}
\usepackage{txfonts}
\usepackage{graphicx}
\usepackage{amsmath}
\usepackage{natbib}
\usepackage{amssymb}
\bibpunct{(}{)}{;}{a}{}{,} 
\begin{document}

\title{The first stars: CEMP--no stars and  signatures of spinstars}
\titlerunning{CEMP--no stars}

\author{Andr\'e Maeder$^1$, Georges Meynet$^1$, Cristina Chiappini$^2$}
\authorrunning{Maeder, Meynet and Chiappini}

\institute{$^1$Geneva Observatory, Geneva University, CH--1290 Sauverny, Switzerland,\\
$^2$Leibniz-Institut für Astrophysik Potsdam, An der Sternwarte 16, 14482, Potsdam, Germany\\
email: andre.maeder@unige.ch,  georges.meynet@unige.ch, cristina.chiappini@aip.de}

\date{Received  / Accepted }

\offprints{Andr\'e Maeder}

\abstract
{}
{The CEMP--no stars are  "carbon-enhanced-metal-poor" stars that  in principle show no evidence  of s-- and r--elements from neutron captures.
We try to understand the origin and nucleosynthetic site of their peculiar CNO, Ne--Na, and Mg--Al abundances. }
{We compare the observed abundances to the nucleosynthetic predictions of AGB models and of models of rotating massive stars with
internal mixing and mass loss. We also analyze the different behaviors of $\alpha$-- and CNO--elements, as well the abundances of
elements involved in the Ne--Na and Mg--Al cycles.   }  
{We show that CEMP-no stars exhibit products
of He--burning that have gone  through  partial mixing and  processing by the CNO cycle, producing low  $^{12}$C/$^{13}$C  and a broad variety of  [C/N] and [O/N] ratios. 
 From a  $^{12}$C/$^{13}$C vs. [C/N] diagram, we conclude that neither the yields of AGB stars (in binaries or not) nor the yields of classic supernovae can fully account for the observed CNO abundances in CEMP-no stars. Better agreement is obtained once the chemical contribution by stellar winds of fast-rotating massive stars is taken into account, where partial mixing takes place, leading to various amounts of CNO  being ejected.
   
The [(C+N+O)/H] ratios of CEMP--no  stars vary linearly with [Fe/H]  above  [Fe/H]=-4.0  indicating primary behavior by (C+N+O). Below [Fe/H]=-4.0, [(C+N+O)/H]   is almost constant as a function of [Fe/H], implying very high [(C+N+O)/Fe] ratios up to 4 dex. In view of the timescales, such abundance ratios reflect more individual nucleosynthetic properties, rather than an average chemical evolution.  The
high  [(C+N+O)/Fe] ratios (as well as the high [(C+N+O)/$\alpha$--elements]) imply 
that stellar winds from partially mixed stars were the main source  of these excesses of heavy elements now observed in CEMP--no stars.
The  ranges covered by the variations of [Na/Fe],
[Mg/Fe], and [Al/Fe] are much broader than for the $\alpha$--elements
(with an atomic mass number above 24) and are  comparable to the wide ranges covered by the  CNO elements. Nevertheless, the ratios [Na/N] and [Mg/Al] are about constant for CEMP--no stars of different [Fe/H]. This is consistent with the view  that the Ne--Na and Mg--Al cycles were significantly operating in the source stars. The very different properties of CNO, Ne--Na, and Mg--Al elements from those of $\alpha$--elements
  further support the idea that these elements (which all give to CEMP--no stars th eir peculiarities) originate   in slow stellar winds of massive stars experiencing partial mixing.}
{CEMP--no stars present a wide variety in the [C/Fe], [N/Fe], [O/Fe], [Na/Fe], [Mg/Fe], [Al/Fe], and [Sr/Fe] ratios. We show that back-and-forth,
partial mixing between the He-- and H--regions may account for this variety. Some
s--elements, mainly of the first peak, may even be produced by these processes in a small fraction of the CEMP--no stars. We propose a classification scheme for the CEMP--no and low--s stars, based on the changes in composition 
produced by these successive back-and-forth mixing motions. }



\maketitle
 
 \section{Introduction}

The  stars belonging to the  first  stellar generations in the Universe, as shown by their very  low metallicity content, had very different physical  properties from  the stars presently forming in the  Milky Way. 
They were much more compact objects, which are denser and hotter than stars with solar composition, and they have a different evolution as well. In particular, even moderately rotating stars reach the break--up limit during the main--sequence (MS) phase. They evolve rapidly to the red after the core H--burning phase (an effect that depends on the treatment of shear mixing),  and important surface enrichment in CNO elements occurs at the supergiant stage \citep{MMEZZEROSTARS,Ekstrom2008,Georgy2013}. 
Some traces of the properties of these first stars may be revealed by the initial chemical enrichment of the Galaxy, owing to their peculiar  chemical yields.

Among the stars with low Fe--content, typically [Fe/H] $<  - 2.5$  (i.e., a number ratio of iron to hydrogen atoms at least 300 times  less than the solar ratio), the  "CEMP--no" stars form an interesting group that dominates at the lowest  [Fe/H] ratios
 \citep{Aoki2002,Masseron2010,Allen2012,Norris3,Norris4}. They may show impressive enhancements,
up to more than a factor $10^4$, of  [C/Fe],  as well as of the [N/Fe] and [O/Fe] ratios. Their distinctive property, which differentiates them
from the other classes of 
CEMP  (carbon-enhanced metal-poor) stars, is the relative absence of "s" and  "r elements", in particular Ba and Eu  synthesized via neutron captures. This is why these stars are called "CEMP--no" stars.  For the definition and properties of CEMP stars, see the review by \citet{Beers2005}.

 CEMP--no stars are  a kind of still-living stellar fossil. They are very old low-mass stars (about 0.8 M$_{\odot}$ or
less), which are surviving from the early times, and they preserve the characteristics of the nucleosynthetic enrichments 
of the early days. 
 Some of these stars are still
on the MS or are subgiants close to it, while others are in the red giant phase. Thus, the particular abundances of CEMP--no stars are not due to
self--enrichment, but are more likely due to these stars having been formed in an environment polluted by the particular nucleosynthetic enrichments from objects called the source stars.

CEMP--no  stars are generally thought to belong to the outer halo population
\citep{Carollo2012} and have also been found in dwarf spheroidal galaxies  \citep{Gilmore2013}. The frequency of CEMP--no stars relative to low-metallicity stars without C-excesses rapidly increases 
for the lowest  [Fe/H] values \citep{Norris4}.  The reason for
the dominant presence of CEMP stars (of various types) at the
low [Fe/H] was interpreted \citep{Gilmore2013} as due to a rapid
gas cooling for a C--rich medium, thus these C--rich stars formed more rapidly than other stars.

 Several  star models and scenarios have been developed to explain the peculiarities of the CEMP stars. They have been  recently 
 reviewed  by  \citet{Nomotoaraa2013}, and are the following:
\begin{itemize}
 \item The models of faint supernovae from Pop. III stars with mixing and fallback \citep{Tominaga2014,Takahashi2014},
\item A two-step scenario with the combination of  a normal supernova  from a 15 M$_{\odot}$ star and a dark supernova from a more massive star (35 M$_{\odot}$) with strong fallback ejecting only its outer layers \citep{Limongi2003},
 \item  The mass transfer from an AGB binary companion that may produce a star with strong CNO enhancements since the AGB yields show these characteristics \citep{Herwig2004,Karakas2007}, 
\item The models of low-metallicity 
stars with an average rotation experiencing efficient rotational
mixing  and mass loss \citep{MMEZZEROSTARS,Meynet2006,Meynet2010}, 
\item  The self enrichment within the observed stars as has been proposed to account for the CN excesses \citep{Campbell2010}.  
\end{itemize}
This last scenario is unlikely,
since  many CEMP stars,  with enhanced neutron capture elements, are still  MS stars \citep{Allen2012}.  An alternative model  
 has also been proposed where  the abundance anomalies of CEMP stars are interpreted as due to "the separation of gas and dust beyond the stellar surface, followed by the accretion of dust-depleted gas" \citep{VennLambert2008}.
This model faces the difficulty of accounting for the low  $^{12}$C/$^{13}$C  ratio found in many  CEMP stars \citep{Masseron2010}.

The scenario of binary mass transfer from an AGB companion appears to  fit many properties of the CEMP--s stars remarkably. For them,
the  enrichments may be due to the $^{13}$C neutron source in low-mass AGB stars, as suggested, for example,  by \citet{Masseron2010}, who provide
several pieces of evidence for that. These authors interpret  
the r-- and s--enrichments in  CEMP--rs stars as resulting
from a unique process: the $^{22}$Ne source during the very hot conditions of the thermal pulses in an AGB star, which then contaminates its low-mass
companion. The binary scenario is also considerably reinforced, at least for the CEMP--s stars, because 
a study of their radial velocities suggests  that all CEMP--s stars are binaries \citep{Lucatello2005}.

The situation is not the same for the CEMP--no stars, for which the binary scenario seems difficult to accept.
 From a study of the binary properties of the different classes of CEMP stars, \citet{Starkenburg2014} conclude that the CEMP-no dataset is inconsistent with the binary properties of the
CEMP-s class. Although the CEMP-no binary fraction is still poorly constrained, their population resembles the binary properties in the solar neighborhood
more, so that their chemical peculiarities are most probably not related to their being  in binary systems.
 \citet{Lee2014}   point out that the relative number frequency of CEMP stars with [Fe/H] < -3.0 (essentially CEMP--no stars) is much too high to be accounted for by the AGB model, and they support the view that "one or more additional mechanisms, not associated with AGB stars, are required to produce carbon-rich material below [Fe/H] = -3.0 ". Also, \citet{Norris4}
carefully examined the possible variations in radial velocities for these stars and found no evidence of variations at the level of  3 km/s in the large majority of them.

The aim of the present work is to examine various abundance ratios of CEMP--no stars and compare them to nucleosynthetic 
properties of massive stars at low metallicities, as predicted by models of rotating stars.
In Sect.  \ref{firstsix}, we briefly review the known properties in the chemical evolution of galaxies that speak for the significant role of mixing and mass loss in the early galactic evolution. In  Sect. \ref{mixingfallback}, we discuss some  properties of the models of rotating massive stars and   of  the "fallback and mixing" models
\citep{Nomotoaraa2013}.  
In Sect. \ref{CEMPCNO}, we show more new evidence or signatures of partial mixing and of mass loss of CNO elements by massive stars
provided by the CEMP--no stars. 
  Section \ref{ca} 
examines the relation or, more exactly, the absence of relation between CNO-- and  $\alpha$--elements  and the constraints that it provides. 
 In Sect.  \ref{nenamgal}, the abundances related to the   Ne--Na and Mg--Al cycles are studied and the differences with the anticorrelations 
found in globular clusters are elucidated. In Sect. \ref{organize}, a  classification  scheme of CEMP--no stars is proposed, and the case of stars with a significant Sr content  is examined.
Section~\ref{concl} provides the
conclusions.

\section{The six known signatures of spinstars}   \label{firstsix}

\subsection{Evolution of the chemical abundances in spinstars}   \label{chem}

The first indication of the possible faster rotation of stars with lower metallicity $Z$ was finding that the relative frequency of Be stars was 
strongly increasing from samples in  the solar neighborhood, to the
Large Magellanic Cloud (LMC) and Small Magellanic Cloud (SMC) \citep{MaederGrebelM}.
This  has been  confirmed by subsequent works \citep{Martayan2007}. It does not mean that this trend is necessarily going on at lower
 $Z$. However, theoretical models of the formation of the first stars indicate 
that  stars of very low $Z$ should have very  high rotational velocities, and thus experience mixing \citep{Stacy2011}. 
The model of spinstars
has been developed  to represent this kind of object \citep{Meynet2010,MM2012Rev}: massive stars generally of low $Z$ with fast rotation,
strong mixing, and high mass loss.

The  mixing in spinstars, due mainly to shear instabilities \citep{maederlivre09}, currently brings products of the CNO-burning   ($^{14}$N and $^{13}$C) to the stellar surface. It may  happen that
 products of the He-burning (C and O)  are mixed  in H-burning regions,  thus  producing  primary $^{14}$N and $^{13}$C \citep{MM2002Nitrogen}, which may reach the stellar surface
(primary means produced from the initial H and He). During H--burning, especially in low $Z$ models, which  are hotter and denser than models at solar composition, the  Ne--Na and Mg--Al reactions (or cycles) also significantly occur that involve the various isotopes of these elements.

In the He--burning regions,  $^{22}$Ne is produced  from $\alpha$-captures on  $^{14}$N.  Further  reactions may also occur  in models of very low $Z$,   such as the $^{22}Ne(\alpha, \gamma)^{26}Mg$ and $^{22}Ne(\alpha, n)^{25}Mg$ reactions, which produce other
daughters of nitrogen. The $^{16}O(\alpha,\gamma)^{20}Ne$ and the $^{20}Ne(\alpha,\gamma)^{24}Mg$ reactions  may also  operate during He-burning in the hot conditions of low $Z$ stars. These various reactions produce elements able to reach the stellar surface and modify the compositions of the winds and the chemical yields of the first stellar generations \citep{Meynet2006}. 

The mixing processes enrich the stellar surface in elements, which like carbon, increase the opacity of the outer layers and enhance the mass loss
 rates of the models having initially very low $Z$. Without mixing, the models would have kept inefficient stellar winds, while with metallicity enrichment the stellar winds can reach  the strength of those in the  Magellanic Clouds \citep{Meynet2006,Hirschi2007}. The consequence is that the contribution of the stellar winds to the chemical yields can be very significant in advanced evolutionary phases. We could even think that the winds may be the main
 source of chemical enrichment for massive stars (say above 25 M$_{\odot}$), which end their life as a black hole with a large fallback. 

 Stellar winds, particularly from red supergiants, have much lower velocities than the ejecta from supernovae.  Thus, the nucleosynthetic production of supernovae is more likely to
  escape  from the region of star formation, while the slower ejecta  by stellar winds could remain kept inside this region.
This possibility is all the more likely in the outer galactic halo to which CEMP--no stars  are thought to mostly belong.
  At the same time, this could also account for the very small enrichments in $\alpha$--elements
with atomic mass higher than that of Mg and other heavy elements observed  in CEMP--no stars. 
This  makes it plausible that CEMP-no stars were formed mostly  from  material coming from the winds of spinstars.

\subsection{First two signatures: galactic chemical evolution of N/O and C/O at low Z}

In a recent review paper, \citet{Chiappini2013} recalled five existing results in the chemical evolution of the Milky Way, which may support the model of spinstars  
acting in the early evolution of galaxies. 
We briefly  recall these five signatures. 

The production of primary nitrogen by rotational mixing in low-metallicity AGB and  massive stars is a significant effect \citep{MMVIII,MM2002Nitrogen,Hirschi2007}. 
These model predictions are in good agreement with the constant N/O ratios observed in very metal-poor normal stars of the galactic halo \citep{Chiappini2006}. Considerations of the lifetimes of galactic chemical evolution   support the view that the  chemical enrichments  are more likely due to massive stars than to AGB stars.
The same kind of result is also present for the C/O ratios. The above models indicate increasing stellar yields of C at lower metallicities,  because more C  is ejected by mass loss (since the surface enrichments are stronger at lower $Z$) and
escapes further nuclear destruction in the star.  These yields lead to good agreement with the observed increase in the C/O ratios
in very low-metallicity stars \citep{Chiappini2006,Fabbian2009}.

\subsection{The third signature:  chemical evolution of $^{12}C/^{13}C$}

\citet{Chiappini2008} have also considered the effects of
spinstars on the  evolution of the ratio $^{12}$C/$^{13}$C. The  predictions of chemical models of galactic evolution 
(without the yields of fast rotators)  indicate values  of  $^{12}$C/$^{13}$C between 4500 and 31000 for 
[Fe/H] ratios between -3.5 and -5.0, while  spinstars lead to ratios between 30 and 300. 
The isotopic ratios observed for very
metal-poor "unmixed" giants \citep{Spite2006} in the galactic
halo with [Fe/ H] between -2.6 and -4.0 lie between 10 and 100.
If the stars observed by Spite et al.  are really unmixed and  have
a surface composition representative of their initial composition, then these observations 
favor spinstars \citep{Chiappini2008}. However, it is not guaranteed that these stars have not suffered the first dredge-up.

\subsection{The fourth signature: primary behavior of Be and B}

A fourth signature of the action of spinstars has been provided
by the study of the formation of the light elements Be and B
(Prantzos 2012).  For some time, it had been shown that B and
Be are produced by spallation of CNO nuclei by galactic cosmic
rays (GCR) in the interstellar medium. The discovery that  Be
and B behave as primary elements  ({\emph{i.e.,}} growing linearly with
[O/H] or [Fe/H]) by \citet{Gilmore1992} contradicted theoretical expectations at that time. Indeed, if GCR protons
hit O nuclei in the interstellar medium, [Be/H] and
[B/H] would grow like 2 [O/H] (secondary elements). The reason for this dependence like  2 [O/H] is that the rate of supernova explosions intervenes  twice \citep{Duncan1992}, because  both 
GCR protons  and oxygen nuclei are produced by different supernovae in the course of the evolution of galaxies.

However, if GCR are accelerated from the CNO rich stellar winds
of rotating massive stars by the forward shock of the subsequent SN, the CNO content in the GCR remains constant with respect to the metallicity \citep{Prantzos2012}. Be or B produced by 
fast CNO nuclei hitting protons and $\alpha$-particles of the ISM will no longer depend on the metallicity, and thus
one expects that the Be/O ratio remains constant as a function of [O/H].

Remarks on the various possible dependences (such as [O/H]
or 2[O/H]) had already been made by \citet{Gilmore1992} and
\citet{Duncan1992}. With models of galactic evolution, Prantzos
studied the observations of the Be and B abundances as a
function of [Fe/H] and emphasized the need for a CNO contribution
by the winds of massive rotating stars to the early chemical
evolution of galaxies.

\subsection{The fifth signature: production of s-elements} 

Another interesting  test has been shown by  \citet{Chiappini2011}. Below [Fe/H] = -2, the ratio  [Sr/Ba] shows an enormous scatter over about  4 dex, extending from [Sr/Ba] = - 2 to about + 2.  We recall here that Sr is the typical element
of the first peak of the s elements, while Ba is the typical representative of the second peak. Thus,  [Sr/Ba]
expresses the ratio of abundances in the first peak with respect to the abundances in the second one.
The standard models for producing r-elements shows a very low scatter
of the [Sr/Ba] ratio  (about 0.2 dex) according to  \citet{Cescutti2013}. The models of binary mass transfer from an AGB star to a lower mass companion \citep{Herwig2004, Karakas2007} account for the stars  very well with [Sr/Ba] between -2 and 0, as confirmed by several authors \citep{Masseron2010,Allen2012}. This is the domain of CEMP-s stars, which in addition are all likely to be binary stars
\citep{Lucatello2005}.

The problem was really to explain the stars with [Sr/Ba] > 0. \citet{Cescutti2013} have shown that chemical evolution models, which also include 
the yields in s-elements produced by fast rotating massive stars,  help to reproduce the distribution of the ratios [Sr/Ba]  toward high values. This is based on the models of rotating massive stars by \citet{Fris2012}, who have shown that massive stars may in some cases moderately contribute to the first peak of s-elements and contribute little to the second
peak and nothing to the third one. Thus, when both the contribution of AGB binaries and that of rotating massive stars  are accounted for, the full extension 
of the observed range of [Sr/Ba] values from -2 to +2 may be  reproduced \citep{Cescutti2013}.

\subsection{A possible sixth signature: the high initial He of a fraction of the stars in globular clusters}

A double MS band has been found first in the 
globular cluster $\omega$ Cen \citep{Bedin2004} and then in several other globular clusters with
even a double sequence of blue stragglers \citep{Dalessandro2013}. The  double MS band was interpreted 
\citep{Piotto2005} as resulting from two populations of stars with different He contents within 
the globular cluster. The bluer sequence has a helium content $Y=0.38$  compared to the standard value of $Y=0.24$. This implies  a He difference $\Delta Y = 0.14$ for a metallicity difference $\Delta Z = 0.002$. Such  He contents  have been confirmed by
study of red giant spectra, which show two different He abundances, one  $Y=0.22,$ and one in the range 0.39 - 0.44, implying a value of at least  $\Delta Y= 0.17$  \citep{DupreeAv2013}. 

The above observations support relative helium to metal enrichments $\Delta Y/\Delta Z$ of about 70 or 
more, while the ratio from standard supernova nucleosynthesis is around 4.  \citet{MM2006omegaC} 
  suggested that this extreme $\Delta Y/\Delta Z$  ratio results from the winds of massive and intermediate stars. However,  we noted
that this is  not the only possibility. It could also result from the escape of the high velocity and high $Z$ ejecta of supernova explosions  from the globular clusters. Models of chemical evolution support the idea that there is an escape of heavy elements in the supernova winds, while the slower He-rich winds of 
massive stars and AGB stars are kept in the shallow potential well of globular clusters  \citep{Romano2010}.
The question as to whether spinstars are involved  in the high He contents observed is not definitely settled.  However the very high observed amplitude $\Delta Y$ in some clusters pleads for huge amounts of He ejected at some stages, and this is not inconsistent
with the role of spinstars in the early galactic evolution.
 A major issue in this context is also to know whether CEMP--no stars  have a high He content as suggested by \citet{Meynet2010}.

In globular clusters,  there are some anticorrelations of chemical abundances  (Sect. \ref{nenamgal})   that point toward H--burning products,
while CEMP--no stars are characterized by both H-- and He--burning products. It is  possible   that the main mass loss phase  
does not occur at the same  stage of stellar evolution.
To account for the anticorrelations  in globular clusters, stellar  mass loss needs to occur during  the core H-burning phase, while 
for  CEMP--no stars, it is while He-burning sources are active.
For globular clusters, reaching the critical limit or close binary evolution with tidal mixing during the MS phase offers interesting candidate effects.
For CEMP--no stars, in contrast, the source stars  need to lose mass in the advanced phases, when both H-- and He--burning products appear at the stellar surface.

One can  wonder what makes this difference in the time 
of mass loss. 
We note  that the metallicity is different, since
CEMP--no stars do occur at much lower metallicities than  globular clusters. This appears to be the main difference.
At a very low metallicity, little mass is lost by a star reaching the critical velocity limit. A lot of mass could be lost  by stellar winds in the advanced phases, when due to self surface enrichment in CNO elements, the opacity of the outer layers become greater and line-driven stellar winds stronger. At a slightly higher metallicity in globular clusters,  stars reaching  the critical rotational velocity could experience strong mass loss \citep{Decressin2007}.
We emphasize that  this remains very speculative. The explanation we have tried to propose may, however, make sense for these 
intriguing differences.

\vspace{1mm}
At this  stage, the six above facts already form an interesting basis for supporting the view that spinstars, with mixing and mass loss,
significantly contribute to the early chemical enrichment of the galaxies, in particular by their large He and CNO yields. Below,
we examine how the properties of the CEMP--no stars further support this view.
 
\section{Comparison between "mixing and fallback" and "spinstar" models}  \label{mixingfallback}

The models of binary mass transfer from an AGB companion  applies well to CEMP-s stars, as seen above. For the CEMP--no
stars, the models of mixing and fallback  and those of massive rotating stars (with mixing and mass loss) are the most
promising according to \citet{Norris4}. Although very different in their physics, they lead
to rather similar consequences, as far as the chemical yields are concerned.
Here, we briefly summarize the main properties of both models. 

\subsection{Models of  "mixing and fallback"}
The models of "mixing and fallback" have been proposed by Nomoto and colleagues \citep{Nomotoaraa2013} to account  for the possible internal mixing in stars and for the fact that only  a fraction of the
"onion skin layers" of the presupernova is ejected. The rest  collapses  into the remnant object, whether  a black hole or a neutron star. 
In mixing and fallback models, the region between two limiting shells, chosen differently in each star model to explain the
observations, is considered to be fully mixed.
The models assume that a fraction of the star is
ejected and another fraction is locked into the remnant. 
The separation essentially occurs at the time of
the supernova (SN) explosion. 
Below some chosen cutoff–mass,
matter is kept in the remnant, and above it matter is ejected and participates
in the further chemical evolution of galaxies. Such events produce faint supernovae   with low kinetic  energy resulting in  a very small amount of $^{56}$Ni ejected  (typically a few $10^{-3}$ M$_{\odot}$)
 \citep{Nomotoaraa2013}. They occur in the range of 10 to 13 M$_{\odot}$, and  up to about 30 M$_{\odot}$ for stars that experience a significant fallback, thus letting only small amounts of heavy elements escape into the interstellar medium.

\citet{Tominaga2014} have performed extensive and impressive comparisons between predicted
values and observed abundances in metal poor stars. They use  models from \citet{Iwamoto2005} for 25 and 40 M$_{\odot}$ without enhanced mixing and a model
of 25  M$_{\odot}$ with enhanced  mixing. The mixing  is made at the time of the SN explosion and occurs between two arbitrary limits. The authors  note that the mixing  could, for example, be due to rapid rotation, but no calculations of rotational effects are performed.
Mixing and fallback in models of faint supernovae
lead to generally good agreement with observed abundances
of CEMP stars as shown by Tominaga et al.
The authors then exploit the parameter space (assuming different values for the region, which is mixed, and for the cutoff mass) for each CEMP star in order to find the best model that matches the observations of that particular object.
This gives much freedom in the adjustments to the observations.

\subsection{Models of "spinstars"}
In the "spinstar" models, a fraction of the stellar interior is chemically mixed. 
The mixing has  a rotational origin, then is progressive during evolution and  not necessarily complete. Finally,      
it results from calculations of shear mixing and meridional circulation; at
the same time, the transport of angular momentum is accounted
for. This mixing is computed with diffusion coefficients, which are also used for comparisons
 between models and observations for rotating stars in the solar neighborhood
(surface compositions and surface velocities, populations of massive stars as red and blue supergiants, Wolf-Rayet stars). 
Thus, the physics used in  very metal-poor stars is the same as in normal metallicity stars. 
What is different are the consequences of the physics implied by the very low metallicity considered.

\subsection{Similarities and differences}
In spinstar models,  the physics triggering the mixing  is discussed in detail and an explanation is provided for why strong mixing  occurs at very low metallicity.
The main difference is that in mixing and fallback models,
the mixing does not occur progressively during the evolutionary stages, 
but occurs at the time of the explosion and with an efficiency that is a free parameter.
In both sets of models,   a fraction of the star is
ejected, and another fraction is locked into the remnant. In spinstar models, the mass
loss rates by the winds in all stages before the SN are calculated, and
this provides the yields of various elements from the winds. The winds are enhanced  by the rotational enrichment of the surface by CNO elements. 
Interestingly, other processes can trigger strong mass losses in metal-free stars and likely in very metal poor stars, too. Recently, \citet{Moriya2014} have suggested 
that the very massive metal-free stars, which evolve into red supergiants, 
become pulsationally unstable shortly before they explode,  thus experiencing  extreme mass-loss rates despite the tiny metal content of the envelopes.
The ejected matter escapes further nuclear destruction in the star.
This particularly concerns some relatively fragile isotopes, like $^{14}$N and $^{13}$C, produced by the CNO cycle. In spinstar models, the yields of these isotopes are relatively high.

 Further enrichment by the subsequent SN explosion may also be modeled or not in spinstar models. (In some models the "wind" and the "wind+ SN"
contributions are both provided.) In that case, as for mixing
and fallback models, the mass cut is considered as a free
parameter, but chosen to be identical for all models.  Some  fallback processes (and perhaps mixing)   could also occur at the time of the SN explosion in spinstars models. It is still unknown  by how much the cutoff mass, which is a critical parameter, is different in models with rotation. A combination of the  
rotational effects and fallback,  and maybe mixing, at the time of SN explosions may provide the best solution. 
A recent work by \citet{Takahashi2014} shows the interest in this kind of approach.

Below in Sect. 4.5, we examine whether some observations allow us to
distinguish between these two kinds of models. 
The inspection of the figures given by \citet{Tominaga2014} shows that the fallback and mixing models often have difficulty accounting for the high observed N–abundances.
Concerning the ratio  $^{13}$C/$^{12}$C ratio, which as shown below has a  strong discriminating power, we have found no predictions
in the quoted models.
In conclusion,  we point out that the stellar wind and fallback+ mixing models appear more complementary than contradictory.

\section{Signatures  of spinstars from CNO data for CEMP--no stars}   \label{CEMPCNO}

\begin{table*}[t!]  
\vspace*{0mm}
 \caption{Abundance data for the sample stars considered } \label{alphalm}
\begin{center}\scriptsize
\begin{tabular}{cccccccccccccccc}
Star  &  $ T_{\mathrm{eff}}$ & $ \log g $ &  [Fe/H] & A(Li) & $^{12}$C/$^{13}$C  & [C/Fe] & [N/Fe] & [O/Fe]  & [$\frac{C+N+O}{H}$] & [Na/Fe] & [Mg/Fe] &[Al/Fe]   & [Si/Fe]  &   [Ca/Fe]   & ref\\
&   &  & & &   &  & \\
\hline
&   &  & & &   &  & \\
BD+44 493 & 5510 & 3.70  & -3.68  &- & -  & 1.31 &     0.32 & 1.59  & -2.20 & 0.27 & 0.52 &
-0.57   & 0.41 &   0.27  & 1\\
BS 16929-005 & 5229 & 2.61  & -3.34  & - & >7  & 0.99 &  0.32 & -  &  -  & 0.03 & 0.30 &
-0.72  & 0.38  &   0.34   & 1\\

CS22166-016 &  5250 & 2.0 &   -2.40& - & -  & 1.02 & - & -  & - & 0.37 & 0.68 &-   & 0.22  &   0.50   & 2\\
CS22877-001  &  5100 &   2.2 &  -2.72 & <1.2 &  >10 & 1.00 & 0.00 & -  & - & -0.24 & 0.29 & -0.72   & -  &   0.42   & 2,3 \\
CS22878-027 & 6319 & 4.41  & -2.51 & -   &- & 0.86    &  <1.06 &  -   &  - & -0.17 &  -0.11 &
-   & 0.07 &   0.07  & 1\\
CS22885-096  &   5050 & 1.9  &  -3.66 & - & - & 0.60 & - & -  & - & - & 0.52 & -0.78   & 0.44 &   0.28   & 2\\
CS22945-017 &   6400 & 3.80  &  -2.52 & - & 6 & 2.28 & 2.24 & <2.36  & <-0.19 & - & 0.61 &-  & - &      & 2b-,3\\
CS22949-037 & 4958 & 1.84  & -3.97  & - & 4  & 1.06 &  2.16 & 1.98  & -2.11 & 2.10 & 1.38 &
0.02   & 0.77  &   0.39   & 1\\
CS22956-028  &   6700 & 3.50  &  -2.33 & -& 5 & 1.84 & 1.85 & <2.47  &<-0.02 & - & 0.58 &-   & - &   -   & 2b-,3\\
CS22957-027& 5170 & 2.45  & -3.19  & - & 6  & 2.27 &  1.75 & -  & - & - & 0.30 &
-0.10   & -  &   0.45 & 1\\
CS22958-042  & 6250 & 3.5 & -2.85 & -& 9 & 3.15 & 2.15 & 1.35  & -0.17 & 2.85 & 0.32 &-0.85   & 0.15  &  0.36   & 2, 3 \\
CS29498-043 & 4639 & 1.00  & -3.49  & - &  6  & 1.90 &   2.30 & 2.43 &  -1.18 & 1.47 & 1.52 &
0.34   & 0.82 & 0.00 & 1\\
CS29502-092 & 5074 & 2.21  & -2.99  & <1.2 & 20  & 0.96 &  0.81 & 0.75  &  -2.16 & -  & 0.28 &
-0.68  & - &   0.24  & 1, 2\\
CS30314-067  &  4400 & 0.7  &  -2-85 & <0.6 & - & 0.5 & 1.2 & -  & - & -0.08 & 0.42 &-0.10   & 0.80 &   0.22   & 2\\
CS30322-023 &   4100 & -0.30  &  -3.39 & - & 4 & 0.80 & 2.91 & .63 & -1.55 & 1.04& 0.80 &-   & -  &   0.30   & 2b-,3\\
CS31080-095  &   6050 & 4.5  &  -2.85 & 1.73 & >40 & 2.69 & 0.70 & 2.35  & -0.38 & -0.28 & 0.65 &-0.95   & 0.05 &   0.17   & 2b-,3\\
G77-61  &   4000 & 5.05  &  -4.03 & <1 & 5  & 2.6 & 2.6 & -  & - & 0.60 & 0.49 & -   & -  &   0.37  & 2, 3\\

HE0007-1832 &   6515 & 3.8  &  -2.72 & - & - & 2.45 & 1.67 & -  & -& - & 0.79 &   &   &     & 3\\
  HE 0057-5959  &  5257 & 2.65 & -4.08  & - &> 2 & 0.86 &  2.15  & <2.77&   <-1.52 & 1.98  & 0.51 & - &  -   & 0.65 & 1\\  
  HE 0107-5240  &  5100 & 2.20 & -5.54 & <1.12 & >50 & 3.85 &  2.43  &  2.30&   -2.16 & 1.11  & 0.26 & <-0.26 &  <0.32  & 0.12  & 1, 2\\
HE 0146-1548 & 4636 & 0.99  & -3.46  & -& 4  & 0.84 &   -        & <1.63 &  - & 1.17 & 0.87 &
0.14   & 0.50  &  0.22  & 1\\
HE 0557-4840 & 4900 & 2.20  & -4.81  & - &  -  & 1.70 &  <1.00 & 2.30 & <-2.68    & -0.18 & 0.17 &
-0.65   &  -  &   0.17  & 1\\
HE1012-1540 & 5745 & 3.45  & -3.47  &- & -  & 2.22 &  1.25 & 2.25  & -1.26 & 1.93 & 1.85 &
0.65   & 1.07 &   0.70   & 1\\
HE1150-0428 & 5208 & 2.54  & -3.47  & - & 4  & 2.37 &  2.52 & -   &  -  & -  & 0.41 &
 -   &  -  &   1.16  & 1\\
HE1201-1512 & 5725 & 4.67  & -3.89  & - & >20  & 1.37 & <1.26 & <2.64 &  <-1.46  & -0.33 & 0.24 &
-0.73   & -  &   0.06   & 1\\
HE1300+0157& 5529 & 3.25  & -3.75  & 1.06 & >3  & 1.31 &  <0.71 & 1.76  & <-2.13  & -0.02 & 0.33 &
-0.64   & 0.87  &   0.39   & 1, 2\\
HE1300-0641  &   5308 & 2.96  &  -3.14 & -  & - & 1.29 & - & -  & - & - & 0.04 & -1.21   & -  &   0.01   & 2\\
HE1300-2201  &   6332 & 4.64 &  -2.61 & - & - & 1.01 & - & -  & - & -& 0.29 &-0.92   & - &   0.29   & 2\\
HE1327-2326 & 6180 & 3.70  & -5.76  & <0.62 &  >5  & 4.26 &  4.56 & 3.70  &  -1.69 & 2.48  & 1.55 &
1.23   & -  &   0.29  & 1, 2\\
HE1330-0354  &  6257& 4.13  &  -2.29 & - & - & 1.05 & - & -  & - & - & 0.32 & -0.93   & -  &  0.40  & 2\\
HE1410+0213  &   4890 & 2.00  &  -2.52 & - & 3 & 2.33 & 2.94 & 2.56  & 0.03 & - & 0.33 & -   & -  &   -  & 2b-,3\\
HE1419-1324  &   4900 & 1.80  &  -3.05 &  - & 12 & 1.76 & 1.47 & <1.19  &<-1.57 & - & 0.53 & -   & -  &   -  & 2b-,3\\
HE1506-0113 & 5016 & 2.01  & -3.54  & - &  >20  & 1.47 &  0.61 & <2.32  &  < -1.41 & 1.65 & 0.89 &
-0.53   & 0.50  &   0.19   & 1\\
HE2139-5432 & 5416 & 3.04  & -4.02  &  - & >15 & 2.59 &   2.08 & 3.15  &  -1.03 & 2.15 & 1.61 &
0.36  & 1.00  &   -0.02   & 1\\
HE2142-5656 & 4939 & 1.85  & -2.87  & - & -  & 0.95 &   0.54 & -     &  -  & 0.81 & 0.33 &
-0.62  & 0.35  &   0.30   & 1\\
HE2202-4831 & 5331 & 2.95  & -2.78 &- & -  & 2.41   &  -     & -   & - & 1.44 & 0.12 &
-   & - &   0.17  & 1\\
HE2247-7400 & 4929 & 1.56 & -2.87  & - & - & 0.70 &   -     & -    &   - & 0.82 & 0.33 &
-  & 0.80 &   0.43   & 1\\

Segue 1-7 & 4960 & 1.90  & -3.52  & - &>50 & 2.30 &  0.75  & <2.21  &  <-1.31 & 0.53 & 0.94 &
0.23  & 0.80  &   0.84   & 1\\
SMSS0313-6708 &   5125 & 2.3  &  <-7.1 & 0.7  & >4.5 & - & - & -  &<-2.49& - & >3.3 &  &   &   >0.1   & 4\\
 53327-2044-515 & 5703 & 4.68 & -4.05  & - & >2  & 1.35&  -      & <2.81 & - & 0.14 & 0.40 &
-0.17   & - &   0.19   & 1\\
   &   &  & & &   &  & \\
   \hline
\end{tabular}
\end{center}
\vspace*{0mm}
{Ref: 1. Norris et al. (2013); 2. Allen et al. (2012);  3. Masseron et al. (2010); 4. Keller et al. (2014) give values of [Li/H], [C/H], [Mg/H],  and [Ca/H]
and upper limits  in the case of  other  elements for this star.}
\vspace*{0mm}
\end{table*}

As already mentioned, the abundances of CEMP--no stars reflect the chemical yields  of some star(s), the source star(s),  in a previous generation, which may be the first one.
We examine here the properties of CEMP--no stars, typically in the range  of [Fe/H] = -2.5 to -7.1, where they occur, and compare them to model data of AGB
and spinstars, when available. We use the data on spectroscopic determinations of chemical abundances collected by \citet{Masseron2010}, \citet{Allen2012}, and completed by recent observations  by \citet{Norris4}. When  different values of the abundance
parameters are given by these three groups, we use the most recent ones, although the small differences between authors have no consequences
on the results. Moreover, the data tables often provide only upper bounds for the chemical abundances.
\citet{Norris4} point out that their sample contains 30\% of near turnoff or subgiants stars, the other ones being red giants.
Two stars (HE 0107-5240 and HE 1327-2326) in their sample only have upper  limits for Ba, thus  they cannot formally be cataloged as CEMP--no objects, but they do point out that all their other properties correspond to CEMP--no objects. We  include them in our sample (Table 1). We further 
discuss HE 1327-2326 in Sect. \ref{sr} 

We also include four other  stars: CS 22945-017, CS 22956-028, CS 31080-095, and HE 1410+0213,  cataloged as CEMP--no stars by \citet{Masseron2010}  and  as Ba--low (with 0< [Ba/Fe]<1)  by \citet{Allen2012}. These stars are
indicated by  "b-" in the last column of Table 1 and labeled with an additional horizontal bar over the representative points in all figures.
Table 1 also contains  two stars cataloged as low-s stars
by \citet{Masseron2010} and considered as "b-" by  \citet{Allen2012}: HE 1419-1324 and CS 30322-023. They are also indicated as above  in the table and  figures. There is a third low--s star given by \citet{Masseron2010}, HE 1001-0243, while it has been cataloged as Ba-rich 
by \citet{Allen2012}. We do not include this star in the sample.

We also include some data  for the  extremely metal-poor star SMSS 0313-6708 (also noted
SMSS J 031300.36-670839.3)  with [Fe/H]  < -7.1 analyzed  recently by \citet{Keller2014}. These authors give upper limits for 
the abundance ratios with respect to hydrogen, except in the cases of  Li, C, Mg, and Ca, where the values of the ratios to H are provided. An upper limit is also given for [Fe/H]. Thus,  we can  give lower limits  for the ratios [C/Fe], [Mg/Fe], and [Ca/Fe].  The other abundance ratios  remain undefined.
  The ensemble of the considered data are given in Table 1,
where $A(Li) = \log \frac{n(Li)}{n(H)}+12.0 $ with $n(Li)$ the abundance in number and as usual 
[Fe/H]= $\log \frac{n(Fe)}{n(H)}-\log \left(\frac{n(Fe)}{n(H)}\right)_{\odot}$.

\subsection{The case of the Li-poor stars}   \label{Li}

A special case  concerns the stars that are Li-poor, {\it{\emph{i.e.},}}with a Li abundance much lower than the cosmological value of $A(Li)=\log n(Li)/n(H)+12= 2.72$. Li is generally destroyed in stars. Thus, if the matter ejected by a star, either by stellar winds or SN explosion, is diluted with original interstellar gas having more or less the cosmological Li abundance, the Li abundance is raised up
to some fraction of the cosmological value. Thus,   the level of the Li abundance in CEMP stars gives information \citep{Meynet2010} on the importance of the dilution factor (defined as the ratio $M_{ISM}/M_{ejected}$ of the matter from the original interstellar gas
to the matter ejected by the star).
Of course, this is correct as long as Li is not strongly destroyed by internal mixing processes having occurred in
the CEMP-no star itself, a hypothesis that is confirmed in the case of HE 1327-2326 by the computations by \citet{Korn2009} and that may apply to  the  dwarf and  subgiant CEMP--no stars.

Thus, we may expect that the Li-poor near MS stars that have not destroyed  the Li in their outer layers reflect
the original composition of the ejecta with  little dilution remarkably well. This 
may strongly  constrain  the properties of the source star, which
has produced the heavy elements observed.  If we take 
$\log g \geq 3.25$ as a limit for MS stars and subgiants, we are only left with three stars having a low Li content: G 77-61, HE 1300+0157 and the very low [Fe/H] star HE 1327-2326.  
These stars  generally  occupy no peculiar position in the various plots.  Moreover, their number
is too small to allow us to draw conclusions.

\begin{figure*}[t!]
\centering
\includegraphics[width=.95\textwidth]{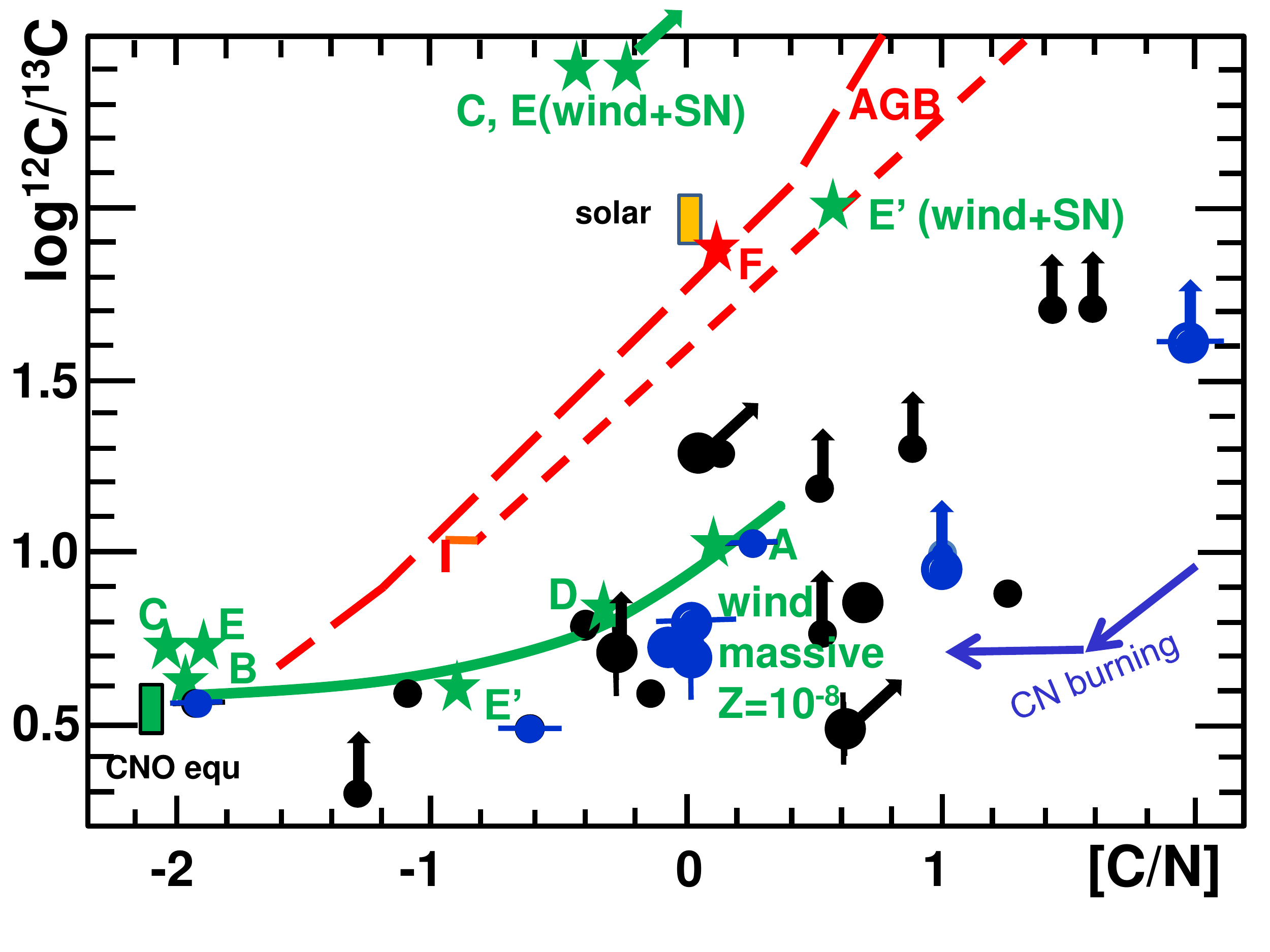}
\caption{Observations of $^{12}$C/$^{13}$C ratios vs.  [C/N],  represented by  black dots \citep{Norris4} and blue dots \citep{Masseron2010}. (The observations by \citet{Allen2012}, represented by black squares in other figures, do not appear in this plot, because these authors 
do not provide data for the C- isotopes.) A vertical  bar indicates MS or subgiant stars with low Li content, and a horizontal bar indicates CEMP--no  or  low--s stars by Masseron et al. which are also cataloged as "b-" by Allen et al.
 The big points apply to MS stars or to subgiants close to the turnoff  with $T_{\mathrm{eff}} > 5500 K$ and $\log g \geq 3.25$,  as by \citet{Norris4},
while the small points represent  the bright giants with lower $T_{\mathrm{eff}}$ and $\log g$. The Li--poor star G 77-61, which  has $T_{\mathrm{eff}}=4000$ K  and $\log g=5.05$, is  considered as a dwarf \citep{Plez2005}, so it is    represented by a big point.
 Some  of the $^{12}$C/$^{13}$C  values are lower bounds indicated by a vertical arrow. If at the same time, the N value is an upper bound, the arrow is oblique.  Rectangles indicate the solar value and the CNO equilibrium value. AGB models with $Z=10^{-4}$ are indicated by broken red lines, the upper line comes from \citet{Karakas2007} in the range of 1 to 6 M$_{\odot}$, and the lower one 
is by \citet{Herwig2004} in the range of 2 to 6 M$_{\odot}$ \citep{Masseron2010}, the low masses being on the top and the high ones at the bottom of the lines. A red star shows a fast-rotating  AGB model of 7 M$_{\odot}$  with $Z=10^{-5}$ (model F) by \citet{Meynet2010}.
Green stars show models of rapidly rotating massive stars  A, B, C, D with masses from 40 to 85 M$_{\odot}$  and $Z=10^{-8}$ from \citet{Hirschi2007}. Model E and E' have a mass of 60 $M_{\odot}$ with  $Z=10^{-5}$ and are given in Table 4 by \citet{Meynet2006},
respectively the left and right columns; they differ by the value of the mass loss rates (higher in E').
A parenthesis (wind+SN) indicates the values when both the wind and the supernova contributions are counted.  
The approximate directions of CN burning is indicated by   blue arrows.
}
\label{C12C13}
\end{figure*}

\subsection{The  $^{12}$C/$^{13}$C   vs. [C/N] relation for CEMP--no stars}  \label{subsc12c13}

The test here concerns CEMP--no stars  in the range  of [Fe/H]= -2.5 to -7.1.
Figure \ref{C12C13} presents the  $^{12}$C/$^{13}$C  vs.   [C/N]  ratios for the stars of Table 1 and compares these data  to those of the  AGB and spinstar models. 
The MS stars  and subgiants  (large symbols) are  distinguished from the bright giants  (small symbols), where mixing may have occurred.
 The many other details concerning this and other figures are given in the caption of Fig. \ref{C12C13}.

 From the physics of nuclear reactions, we may say the following. 
 At the very  beginning of CN burning, $^{12}$C burning produces some  $^{13}$C and $^{14}$N, which  makes some significant
slope in Fig. \ref{C12C13}. Then the ratio $^{12}$C/$^{13}$C  stays constant, while $^{12}$C continues to be turned to $^{14}$N, which produces a flat slope in the figure.
 High ratios $^{12}$C/$^{13}$C  and [C/N] indicate very partial H--burning by the CNO cycle, as well as a possible $^{12}$C addition (for example by a supernova or a WC star). Moderate and  low ratios $^{12}$C/$^{13}$C  (and low [C/N])  cannot be anything but  a signature of H-burning by the CNO cycle, since further  reactions would destroy $^{13}$C by $\alpha$--captures, if $^{13}$C  is brought into a region of He--burning.
Thus, the isotopic   $^{12}$C/$^{13}$C and the [C/N] ratios provide a constraining test on the degree of CNO processing. 
From Fig. \ref{C12C13}, which is  rich in information, we draw the following results:\\

\subsubsection{Sign of partial mixing}
 The solar values, as well as those of CNO equilibrium in massive stars, are indicated.
CEMP-no stars have low $^{12}$C/$^{13}$C ratios providing evidence of  CNO burning. However, the relatively huge amounts of C (as well as O) necessarily imply some  synthesis of C and O by  the 3$\alpha$ reaction during He--burning. Thus, CEMP--no stars generally exhibit products
of He--burning that have gone  through  partial mixing and some  processing by the CNO cycle, producing low  $^{12}$C/$^{13}$C  and a 
broad variety of  [C/N] ratios.

We emphasize   the partial character of the mixing. Partial mixing  means that there are  not a large number of turnovers between the H-- and He--burning regions.
Some C and O coming by mixing  from the He--burning region has been turned into nitrogen, but some variable fractions of this C and O have not been further processed by the CNO cycle.  (This is like a further addition of C and O.)
If we had full mixing as in a convective zone, the $^{13}$C would be rapidly  destroyed by the $^{13}C(\alpha, n)^{16}O$ reaction by going into
the He--burning region,
 leading to high   $^{12}$C/$^{13}$C ratios.
The stars would show s--elements, resulting from neutron captures  of the neutrons emitted by the reactions  
$^{13}C(\alpha, n)^{16}O$ and  $^{22}Ne(\alpha, n)^{25}Mg$.
To put it precisely, CEMP--no stars are generally characterized by an absence of s--elements (but see Sect. 7), thus this is  consistent with partial mixing.

Then, if the matter of the outer stellar layers of the source star is ejected  by stellar winds, it escapes further nuclear processing.  It is thus possible by the combination of 
partial mixing,  CNO processing,  and  mass loss to produce  chemical yields corresponding to the kind of abundances observed in CEMP--no stars.

One  low s--star   (CS 30322-023) shows abundances  corresponding to the CNO equilibrium values. This is an indication of  very  strong CNO processing either in the source star or in the star itself, but  of incomplete  mixing between the H- and He-burning regions, since the amount of s--elements is  low.  Partial mixing is also  consistent with the presence of $^{13}$C in this star. However,  it is the extreme giant
 in the sample, and  there is probably some large CNO processing in this object (as well as some Li destruction).  Nevertheless,  because many MS and subgiant stars
(noted by large points) have very low $^{12}$C/$^{13}$C, we may consider that the  CNO processing is generally coming from the source stars and not from self--enrichment.\\

\subsubsection{Discrepancy with the AGB nucleosynthesis}
 The model data for AGB envelopes with masses from 2 to 6 M$_{\odot}$ and metallicity Z=0.0001  \citep{Herwig2004} and those   of 1 to 6 M$_{\odot}$ and  Z=0.0001   \citep{Karakas2007}  are also shown in Fig. \ref{C12C13}, see also \citet{Masseron2010}.  
  The comparison with observations reveals a  large difference between CEMP--no stars and the location of AGB stars in this diagram, as already shown by \citet{Masseron2010}. For all the [C/N] values, the predicted $^{12}$C/$^{13}$C ratios are generally higher than   observed, even in 
 the region between the solar and the CNO equilibrium values.  The composition  of a 7 M$_{\odot}$ fast-rotating E-AGB envelope 
\citep{Meynet2010} agrees with the other AGB data and thus also with observations. Thus, the yields from AGB models have difficulties to account for the abundances of CEMP--no stars. 

In models of galactic chemical evolution, stars may only contribute to the chemical enrichments after a certain time given by stellar evolution
properties. In the case of AGB stars, their yields influence the chemical abundances   only for [Fe/H] ratios higher than  about -3.0, as shown for example in Fig. 10 by \citet{Nomotoaraa2013}, and also \citet{Chiappini2008} and \citet{Matteucci2012}. In such galactic models for the halo, a Fe/H equal to -3 occurs at an age of 45 Myr, which corresponds to the lifetime of a 7 M$_{\odot}$  star. Therefore AGB stars of moderate masses should not play a significant role for the present plot of CEMP--no stars. \\

\subsubsection{Stellar winds and SN contributions}
Some models \citep{Meynet2006, Meynet2010} provide the total yields, summing the wind and the supernova contributions.  These models are represented in the figure with a parenthesis "(wind+SN)".  These results cannot account for the CEMP-no observed abundances (see also \citet{Cescutti2010}.  We notice that  the lines that would connect models C, E, and E' to models C(wind+SN), E(wind+SN), and E'(wind+SN) have a slope of about 1.0 corresponding to the addition of $^{12}$C. The physical
reason  is clear: supernovae produce much $^{12}$C, while they produce  little $^{13}$C and $^{14}$N, thus making a slope of 1.0. 
This plot indicates  that CEMP--no stars do not receive the complete addition of the    $^{12}$C layer from the source stars from the "onion skin
 model" during supernovae explosions.\\

\subsubsection{Spinstar stellar winds}
Models taking the chemical enrichment due to winds of rotating massive stars in the range of 40 to 120 M$_{\odot}$ and a variety of mass loss rates into account \citep{Meynet2006,Meynet2010} are represented in Fig. \ref{C12C13}. In these models, significant  amounts of C and O produced by He-burning in the core are
transported by mixing processes (mainly shear diffusion) in the H--burning shell. There, the  new C and O may participate in the CNO cycles and be 
more or less transformed to $^{13}$C and $^{14}$N. Depending on whether the burning is complete or incomplete, the fractions of $^{13}$C and $^{14}$N produced may be different, and the same is true for the fractions of $^{12}$C and $^{16}$O. This produces the variety of abundance ratios observed in this diagram (as well as in Fig. \ref{ONCN} below).

The massive star models reproduce the flat part of the curve of the observations   well on the lefthand side  of  Fig. \ref{C12C13}  before  the observed data rise for [C/N] > 0. 
The relatively  flat part on the left results from the fact that    $^{12}$C/$^{13}$C nearly keeps its equilibrium value even when the whole CNO cycle is not at complete equilibrium. On the right of the plot, the CNO burning is very partial, making higher $^{12}$C/$^{13}$C and [C/N] ratios.  

 We notice that the models of spinstars    do not 
completely cover the range of the observed values of CEMP--no stars. They cover the range of abundance ratios from CNO equilibrium values
up to $[C/N]  \approx 0.2$, 
but not the higher [C/N] ratios. In this respect, we emphasize that the initial  values of  [C/N] and [O/N]    play a major role in the location of the model points in Figs. \ref{C12C13} and \ref{ONCN} and may be responsible for some differences. We consider two cases to illustrate this
problem:

A) The case of   mild mixing with a moderate entry of C and O from the He--burning core into an active H--burning zone.  The representative model points
in the figures will be at some distance from the initial values in the direction of the CNO equilibrium point. The size of the deviation from the initial values depends on the amount of mixed C and O. Small entries of C and O make small deviations, large entries make large ones.

B) The case of  mixing with an entry  of C and O into a relatively inactive H--burning zone. The formation of   nitrogen 
will be very small. Thus, the representative model points  are shifted upward and rightward (with high [O/N] and [C/N]) in the mentioned figures with respect to the chosen initial
values. The size of the shifts depends on the amounts of new C and O entered into the  inactive H--burning zone. (In the two cases, the history of 
the mass loss rates of the source star may also play a role.)

In both the  spinstar and AGB models, the solar abundance ratios have been adopted by their authors, in the absence of better information. We suggest that the initial composition had in reality none or very tiny amounts of $^{14}$N, while C and O were nearly scaled on  the (low) abundances of the  other $\alpha$--elements made in supernovae (which is not so much the case for nitrogen). Thus,  the true initial [C/N] and [O/N] ratios were very likely those  about corresponding  to the  stars lying near
 the upper righthand corner of Figs. \ref{C12C13} and \ref{ONCN} (see  also expression (5) below).

For such initial ratios,
 mixing like in case A would distribute the model points from these  high initial ratios all the way down to the left of the figures. That solar ratios have been adopted in spinstar models 
explains why the righthand part of the figures was not covered by the models.

The representative points of some AGB models are located to the right upwards of the solar values in Figs.  \ref{C12C13} and \ref{ONCN}. These are   AGB stars with low initial masses ($\leq$ 2.5 M$_{\odot}$). This location is consistent with the above type B scenario. The H--burning layers in these low-mass AGB stars have relatively low temperatures( see Fig. 26.18  in   \citet{maederlivre09}. Hot bottom burning  is also absent  in these stars (below 4 M$_{\odot}$). As a consequence  of these effects,   the CNO burning remains weak, little nitrogen is produced, and mixing, even if convective, may only lead to largely positive [C/N] and[O/N] values. This explains why some AGB models
have [C/N] and [O/N] higher than solar.

 As a result, solar ratios should not be taken as initial values in future  spinstars and AGB models aimed at being the sources  of CEMP--no stars.\\

\subsubsection{Differences between spinstar and mixing and fallback}
 In light  of
Sect. \ref{mixingfallback}, it is likely that  the predictions of the two sets of models
should be different, although this cannot be quantified yet. The main characteristics of CEMP--no
stars result  mainly from the mixing of the products of
He--burning into the H--burning shell. Thus, 
   mixing at the time of the SN explosion or a progressive and partial mixing during the evolution should lead to significant differences.  The physical conditions and timescales of the two cases are not the same, and this should influence the abundance of  temperature -sensitive  isotopes like  $^{13}$C and $^{14}$N.

In conclusion,  model grids of spinstars with  appropriate initial abundance ratios, a broader range of  rotation velocities and mass loss rates, as well as  predictions  for the $^{12}$C/ $^{13}$C ratios from mixing and fallback models,  are all very  needed in the future to extend the comparisons with observations.

\subsection{The [O/N] vs. [C/N] relation for CEMP--no stars}  \label{subsoncn}

\begin{figure}[t]
\begin{center}
\includegraphics[width=9.1cm, height=7.0cm]{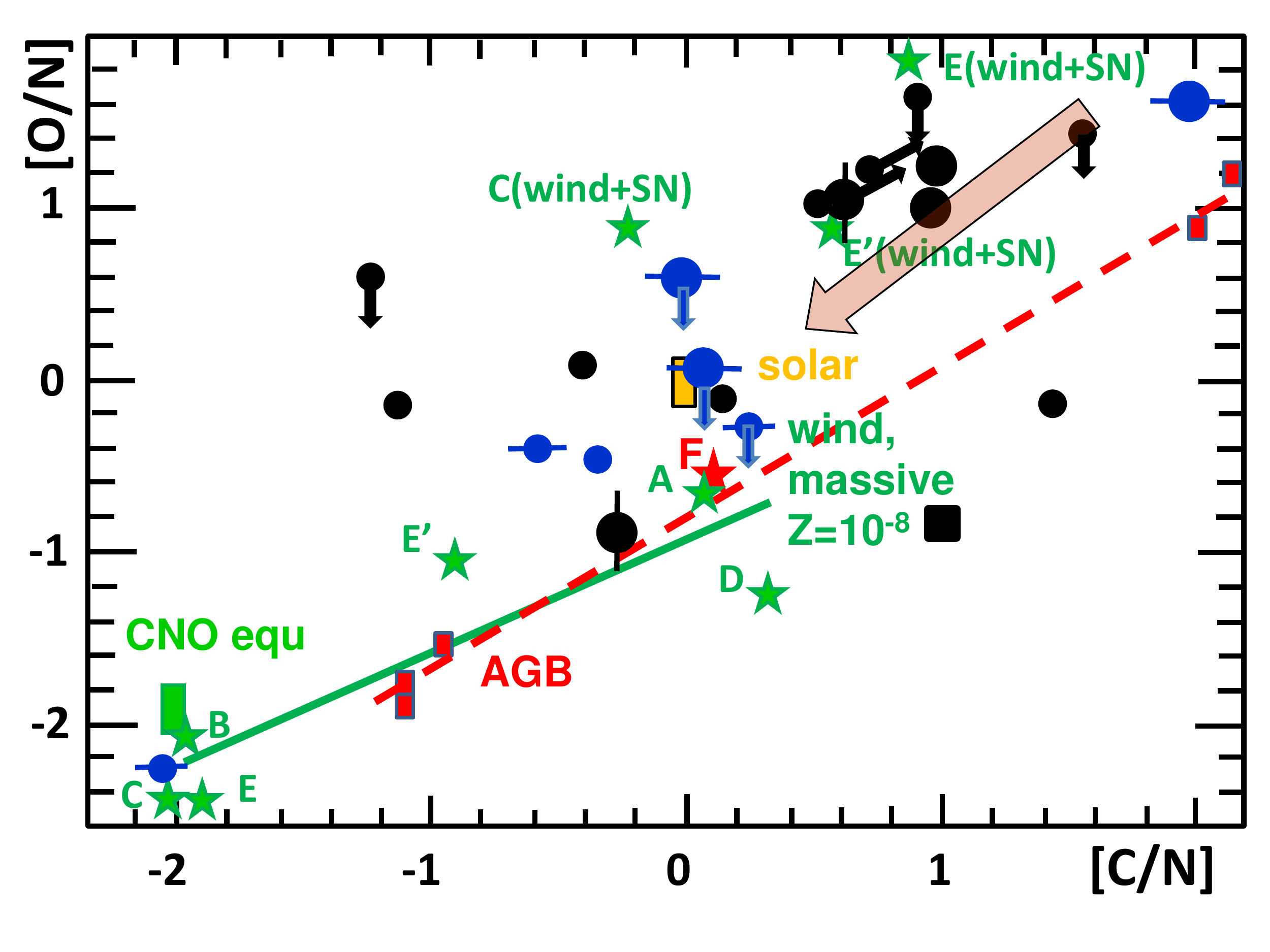}
\caption{Comparison of the abundance ratios [O/N] and [C/N]  of CEMP--no stars with model data.  The same codes as in Fig. \ref{C12C13} 
  are adopted to represent the observations and models. The small red rectangles  show the AGB models by \citet{Herwig2004} connected by a red broken line.
   The  general trend of the CNO burning is represented by a big reddish arrow. The small vertical arrows indicate 
 upper limits for [O/Fe] and the small oblique arrows upper limits for [N/Fe]. The star HE 1201-1512
was not indicated because  of many lower limits.}
\label{ONCN}
\end{center}
\end{figure}

 Figure  \ref{ONCN} shows the [O/N] vs. [C/N] diagram (labels as in Fig. \ref{C12C13}). The stars in this figure are not exactly the same as the ones shown in Fig. \ref{C12C13} because not all stars have  $^{12}$C, $^{13}$C, and oxygen abundances reported.
Most stars in this figure are located in the upper righthand quadrant, hence far from the CNO equilibrium value. This again suggests partial mixing and CNO-processing to have occurred in these objects.

A recent study \citep{Maeder2014} of the N/C vs. N/O plot, or similar plots  like Fig. \ref{ONCN},
shows that the  slope for moderate CNO burning is model independent. For example,  over a range of log (N/O) ratios of about 0.5 dex, the scatter in
log (N/C ) should be less than 0.15 dex, regardless of the models. The curve in this kind of diagram is determined by:
\begin{itemize} 
\item The physics of nuclear processes.
\item The initial number ratios  of CNO elements.
\end{itemize}
 

We examine the physics of this particular plot [O/N] vs. [C/N]. We can make two kinds of approximation. In the first one, we may consider the very beginning of CNO burning (which is also the case of low-mass stars): oxygen Stays constant, while only the CN cycle operates. For simplification,
we call C, N, and O the abundances in numbers. Thus, we have
\begin{eqnarray}
O = const.   \quad \mathrm{and} \quad  dC =  - dN , \\
d(C/N)=\frac{dC}{N} \left(1+ \frac{C}{N}\right) \,  \, \mathrm{and} \, \,  d(O/N)= - \frac{O}{N^2} dN \, .
\end{eqnarray} 
The slope of the relation in Fig. \ref{ONCN} is therefore given by
\begin{eqnarray}
\frac{d \log(O/N)}{d(\log C/N)} \,  = \frac{C}{O} \frac{d(O/N)}{d(C/N)} \,  = \, \frac{C}{N} \, \frac{1} {\left(1+ \frac{C}{N} \right)} ,
\label{ratio1}
\end{eqnarray}
\noindent
where the initial C/N ratios can be used.

We have typical relations between the bracket terms and number ratios, as  in the example below,
\begin{eqnarray}
\log (C/N) = [C/N] + \log (C/N)_{\odot} \, .
\end{eqnarray}
We adopt 
 (C/N)$_{\odot}= 4.03 $  and (O/N)$_{\odot}= 7.59$ \citep{Asplundetal05,Asplundetal09}.
The initial values of (C/N) and  (O/N) are determined by the C and O matter diffused into the H--burning shell. Formally, in the case considered here, these ratios are both
infinite at the beginning of CNO processing, thus Eq. \ref{ratio1} leads to a slope of 1.0 in Fig. \ref{ONCN}.  If we   adopt  the highest  values in the sample stars for the initial values
of 
(C/N)   and (O/N), {\emph{i.e., }}where relatively  less N has been processed,  we have
\begin{eqnarray}
 [C/N]= 2.0 \quad \mathrm{and} \quad  [O/N]=1.6  \, ,
\end{eqnarray}
which lead to
C/N =  403  and  O/N =  302.
The slope of the relation O/N vs. C/N is thus
\begin{eqnarray}
\frac{d(O/N)}{d(C/N)} \,  = \, 0.998 \, \,  , i.e. \, \mathrm{close \, \, to \, \, one.}
\end{eqnarray}

In the second approximation, we  use the fact that in massive stars the CN cycle rapidly brings the C-content to an almost  constant value, so that
the increase in the N--content mainly results in the O--destruction. Thus, we have
\begin{eqnarray}
C = const.   \quad \mathrm{and} \quad  dO =  - dN , \\
d(O/N)=\frac{dO}{N} \left(1+ \frac{O}{N}\right) \, \, \, \mathrm{and} \, \, \,
d(C/N)= - \frac{C}{N^2} dN \, ,\end{eqnarray}
and the slope  becomes
\begin{eqnarray}
\frac{d \log(O/N)}{d(\log C/N)} \,  = \frac{C}{O} \frac{d(O/N)}{d(C/N)} \,  = \, \frac{N}{O} \, \left(1+ \frac{O}{N} \right) . 
\label{ratio2}
\end{eqnarray}
With the number ratios we had above, we get a slope of 1.003. This means that, regardless of  the assumptions,   a slope of  1.0
represents the CNO processing in Fig. \ref{ONCN} well. This slope is represented by a broad arrow in this figure.

From Fig. \ref{ONCN}, we note  the following  points:

1. It is noticeable that CEMP--no stars span a range of more than  $10^3$ in the C/N and O/N ratios!  About 77\% of the CEMP--no stars have either C/N or O/N ratios higher than solar, while currently MS and red giants have values that only cover a fraction of the range between solar and equilibrium values. 
 These high ratios are consistent with  products of He--burning having experienced very partial CNO processing. 

In a He--burning region, both the C/N and O/N ratios tend to infinity, thus the evolution due to the CNO processing moves from the  upper righthand corner  down to the lower lefthand corner in Fig. \ref{ONCN}, as shown by the broad arrow.  These high ratios may also be consistent with the addition of fresh carbon and oxygen from the He-burning
region of the source star.

2. With [O/N] on the vertical axis, there  is now some continuity in models of AGB envelopes and  models of massive rotating stars, which
 was not the case with  $^{12}$C/$^{13}$C on the vertical axis. 
The reason is that the [O/N] ratio does not   reach an equilibrium value early during H-burning and keeps some
intermediate values, while $^{12}$C/$^{13}$C rapidly  comes close to its equilibrium ratio and is a more discriminating parameter, as shown above.

3. The models of AGB and winds of massive stars both have  difficulty reproducing the observed points.
Again, as discussed in the previous section, this may be, at least in part, related to the choice of the initial solar [C/N] and [O/N] ratios in both kinds of models. Indeed, if we adopted the initial values given in relations (5), we see that the observations would lie more or less
on a straight line connecting these initial values to the those of the CNO equilibrium. This also supports the view of partial mixing of He--products
into the H--burning region.

 Comparisons are made here between the compositions of
pure ejecta of one given source star and CEMP-no star surface abundances.  Actually, the CEMP-no star
abundances may result from many more various circumstances;  for instance, it can result from the mixing of the
ejecta of more than one star, some dilution with ISM  also occurs (however, dilution with metal-free
material would have no impact on any ratios of heavy species), and other complexities may intervene to shape  the observed abundances. Another point that remains to be
  examined is the following one: some of the CEMP-no stars are likely He-rich,  with a mass fraction $Y$ of helium at the surface
between 0.30 and 0.60 \citep{Meynet2010}.  This reduces the opacity and modifies the outer stellar structure,
making it denser. One notices that four stars in Table 1 have $\log g > 4.60$, which is surprising for such old objects, and this tends to support the hypothesis of a relatively high density for these objects.
One might wonder whether the  determinations of stellar parameters and abundances are affected  by such possible  high He--contents.

Finally, we  would like to point out that the AGB  and rotating massive star models are often considered to be different (and sometimes opposite) scenarios. This is not very meaningful. For interpreting CEMP--no stars, which typically occur for [Fe/H] <  -2.5,   models of both massive stars  and intermediate-mass stars should be considered down 
to a mass  limit  based on  the timescales  compatible with  galactic evolution models \citep{Matteucci2012}. As mentioned above, for the present sample a lower mass limit of 7 M$_{\odot}$ seems reasonable.

\begin{figure}[t!]
\begin{center}
\includegraphics[width=9.1cm, height=7.0cm]{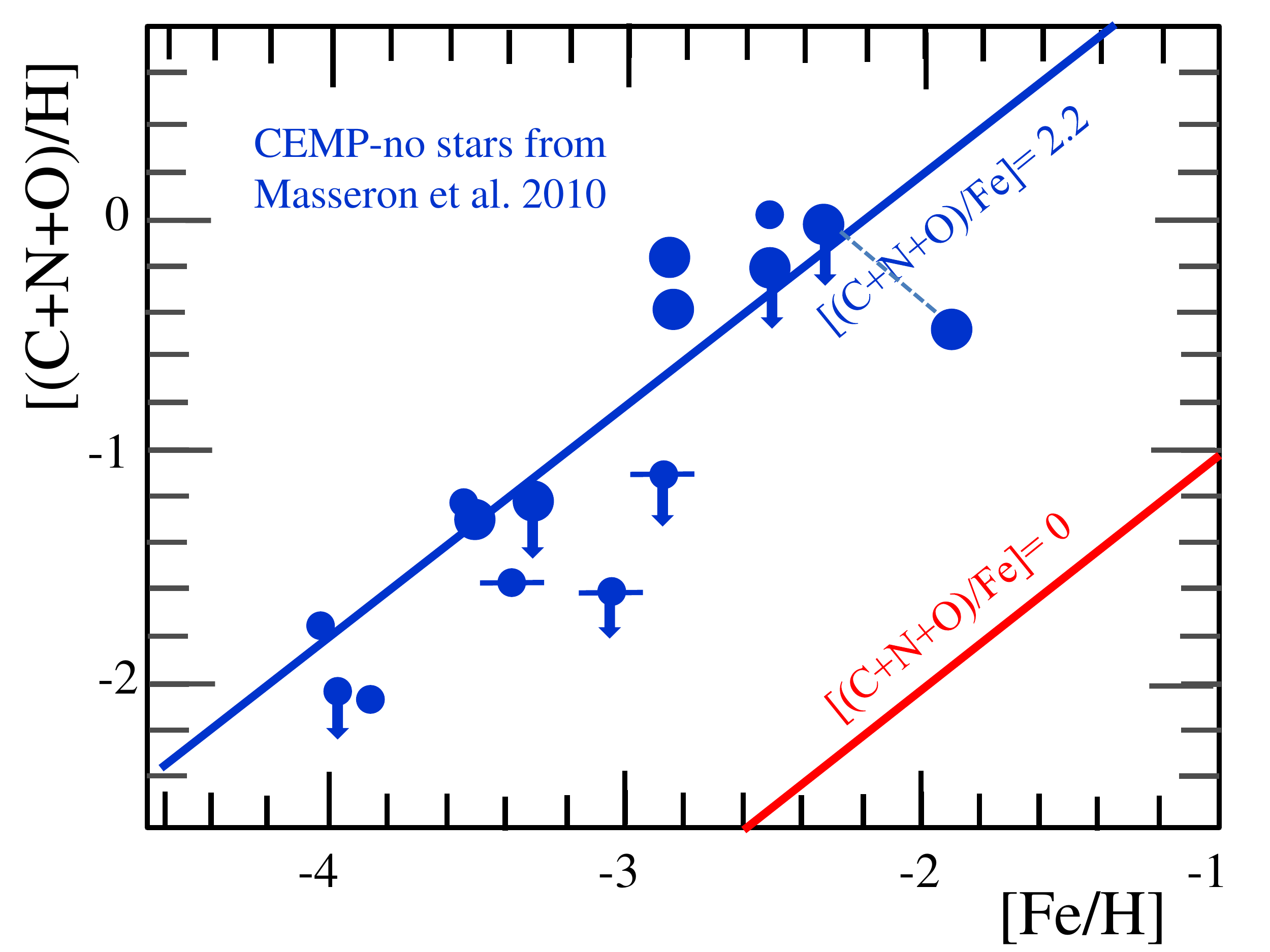} 
\caption{Sum of CNO elements  [(C+N+O)/H] vs. [Fe/H] for CEMP--no stars  with indications of upper limits as in Fig. 21 from \citet{Masseron2010}. The large points represent MS or subgiant stars near the MS turnoff, while smaller points represent giant stars. The small horizontal lines on the circles  denote  low s-elements.  Two lines of constant [(C+N+O)/Fe]  ratios are indicated.}
\label{CNOFE} 
\end{center}
\end{figure}

\subsection{The sum of  CNO elements at extremely low [Fe/H]}  \label{sum}

Masseron et al. (2010) have shown that the sum of CNO elements in CEMP-no stars correlates linearly with metallicity. In Fig.  \ref{CNOFE} we plot the Masseron et al. data along with three other stars showing low abundances of s-elements. This linear relation is not expected according to standard scenarios (see Masseron et al. for a discussion).


In terms of galactic chemical evolution, a slope of 1.0 as a function of metallicity is an indication of primary behavior. This would imply that the sum of CNO elements over Fe remains constant. In the framework of the spinstar scenario, this happens naturally because 
C and O are primary elements made in the He--burning core, and their production depends on the massive star formation rate as for supernovae. Thus, the ratio (C+O)/Fe should be constant. However, because the C and O elements then
undergo  partial H--burning creating some N, this is the sum  (C+N+O)/Fe that stays constant, instead of (C+O)/Fe.

\begin{figure}[t!]
\begin{center}
\includegraphics[width=9.1cm, height=7.0cm]{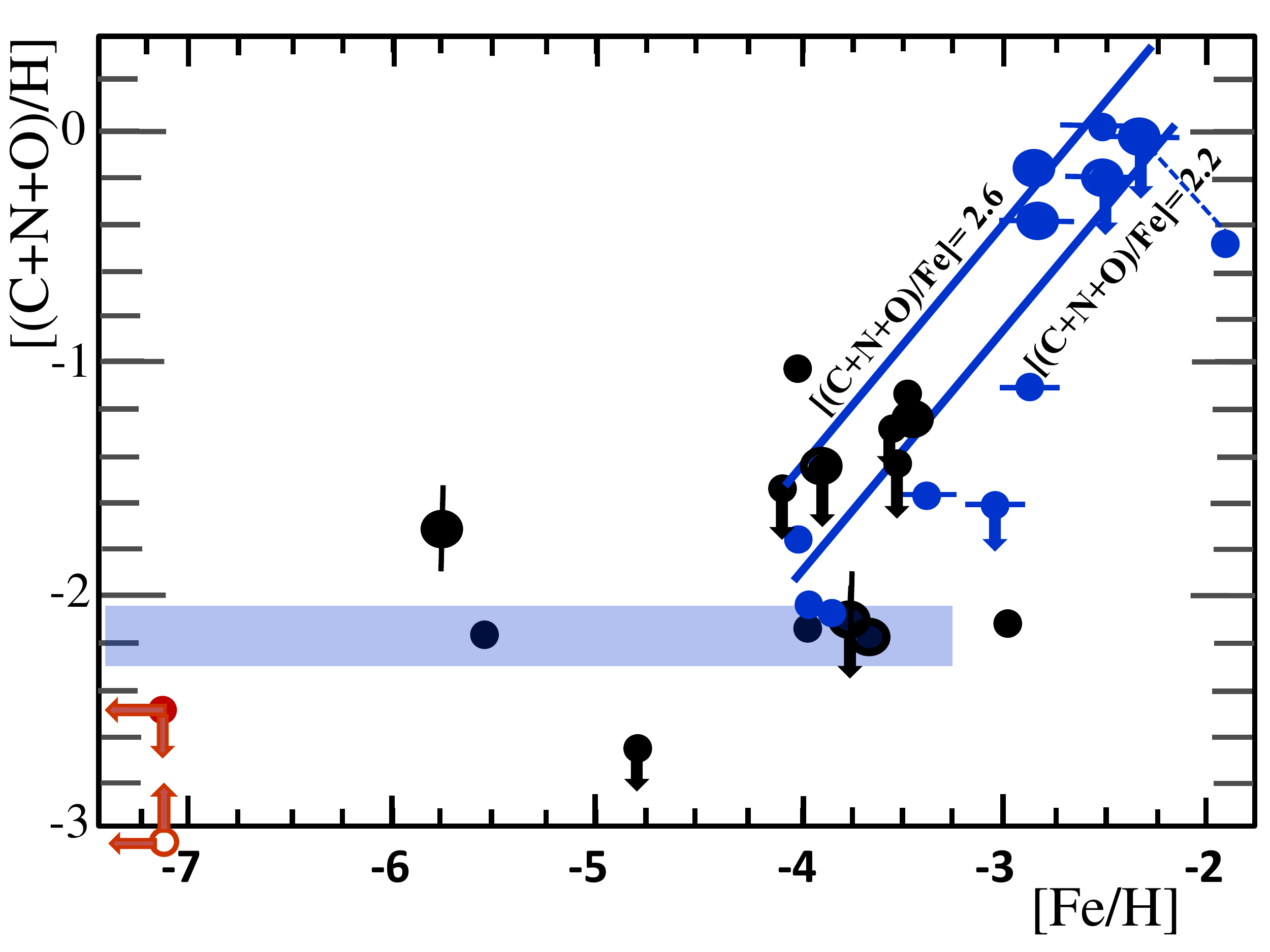} 
\caption{Sum of CNO elements  [(C+N+O)/H] vs. [Fe/H] from various sources down to the lowest [Fe/H] known. The 
codes for representing stars are the same as in Fig. 1.  Upper bounds on   [(C+N+O)/H] are  indicated by arrows.  The brown points at [Fe/H]= -7.1 represent the star SMSS 0313-6708, which has the   limits as  indicated in the text. }
\label{CNOLIPOOR}
\end{center}
\end{figure}

As shown by Fig. \ref{CNOLIPOOR}, the more recent data by \citet{Norris4} and \citet{Keller2014} allow us to extend the plot toward lower [Fe/H] values than in  Fig.~\ref{CNOFE}. We notice several facts:
\begin{itemize}
\item First, the linear relation found by \citet{Masseron2010}  for CEMP--no stars with [Fe/H] > -4.0 is further supported.
\item  Surprisingly, the decreasing linear relation does not go on for lower [Fe/H].
Below a limit around [Fe/H] = -3.5 to -4.0,  the 
extremely iron poor stars keep a more or less constant  [(C+N+O)/H] ratio, or it decreases only very slowly. This means that [(C+N+O)/Fe] becomes very high for very low 
values of [Fe/H].
\item There are  MS stars and subgiants both with high and low   [(C+N+O)/H] ratios,  and the same for the giant stars. Thus, the difference
between the steep and the flat relations is probably not due to self--enrichment  in CEMP--no stars. 
\item The two objects  that are  Li--poor and near MS stars  have relatively  low [(C+N+O)/H] ratios. However, the number of 
such stars is very small.
\item The case of SMSS 0313-6708 with [Fe/H] < -7.1 deserves some explanations.  Mostly upper limits of the abundance ratios [X/H] are given by \citet{Keller2014}. We can  estimate an upper limit [(C+N+O)/H] < -2.49, which is indicated in Fig. \ref{CNOLIPOOR}. Then, if we assume that the sum of C+N+O in the star just consists of C, with no N and O (which is certainly not the case), we get a  lower limit   [(C+N+O)/H] > -3.09. (Here, one has to be careful about the fact that N and O are not zero in the Sun, thus [(C+N+O)/H] $\neq$ [C/H] in this case!) This lower limit is also  indicated in the figure. We see that the [(C+N+O)/H] ratio is relatively well confined.
\item How do we interpret  the very high [(C+N+O)/Fe] ratios up to -4 for the lowest [Fe/H]?
 We first notice that in view of the models of chemical evolution  of the Galaxy, [Fe/H] values below -4 are relevant to very short stellar lifetimes and thus imply source stars with high masses,  above 30 M$_{\odot}$. 
In such short times, it is not possible to build an average chemical galactic evolution, all the more so  for    objects belonging  to the outer galactic halo.  In this most early period, the abundance ratios reflect the individual nucleosynthetic contributions more, rather than an average composition of the galactic environment, hence the need for inhomogeneous chemical evolution models  \citep{Cescutti2014}. 
  Such high [(C+N+O)/Fe] ratios indicate that the typical yields from classical supernovae \citep{Arnett1996,Maeder1992}  do not significantly contribute to the early galactic enrichment of CEMP--no stars,  which may   preferably result from
spinstars, mixing and fallback models, or some combination of both.
Whether the high-velocity SN ejecta escape from the region of further star formation could also be a part of the explanation.
\end{itemize}

Thus, the behavior of the [(C+N+O)/H] provides interesting information on the very early chemical evolution of the Galaxy, showing the level of [Fe/H] above which 
cumulative effects of the chemical enrichment start to be seen.

\subsection{The case of nitrogen}  \label{nitrogen}


Models of supernovae with mixing and fallback, constructed to interpret CEMP--no stars, generally  have  some difficulty producing sufficient nitrogen, and the [C/N] ratios are  often higher by 1 or 2 dex than observed. For example, the comparison of the example  CEMP--no star CS 29498-043
with the appropriate model by \citet{Nomotoaraa2013} (see also \citet{Tominaga2014} shows that the observations  give [C/N] = -0.40 \citep{Norris4}, while the 
SN model with mixing and fallback gives  [C/N] = + 1.6. The models of normal supernovae and hypernovae with mixing and fallback  constructed to interpret stars with
[Fe/H] in the  range of -3.5 to -4.2 also produce insufficient N by about one to  two orders of magnitude.
Thus, nitrogen may be a problem for the mixing and fallback models, which generally  show remarkable agreement
for heavier elements.
The very constraining ratio $^{13}$C/$^{12}$C would further
help to distinguish between the two sets of  models. 

The contribution of the winds of  the "source stars" is likely to be responsible for  the particularities  of CEMP--no stars.
These particularities necessarily originate in massive stars in view of the timescale of galactic chemical evolution. Massive stars become
  red supergiants with high CNO abundances,  a composition known to lead to dust formation and   high mass loss rates  \citep{VanLoon1999,VanLoon2005,Srinivasan2011}. Thus, as stated in Sect. \ref{mixingfallback},  a combination of the
rotational effects (mild mixing and  high mass--loss rates due to composition changes), together with the possible effects of fallback  (and may be further mixing)  at the time of SN explosions, may provide  satisfactory results. A recent work by \citet{Takahashi2014} confirms the  
increase in the abundance of nitrogen (and daughter elements Na and Al) resulting from including the effects of rotational mixing in the models.

\begin{figure}[t]
\begin{center}
\includegraphics[width=9.1cm, height=7.0cm]{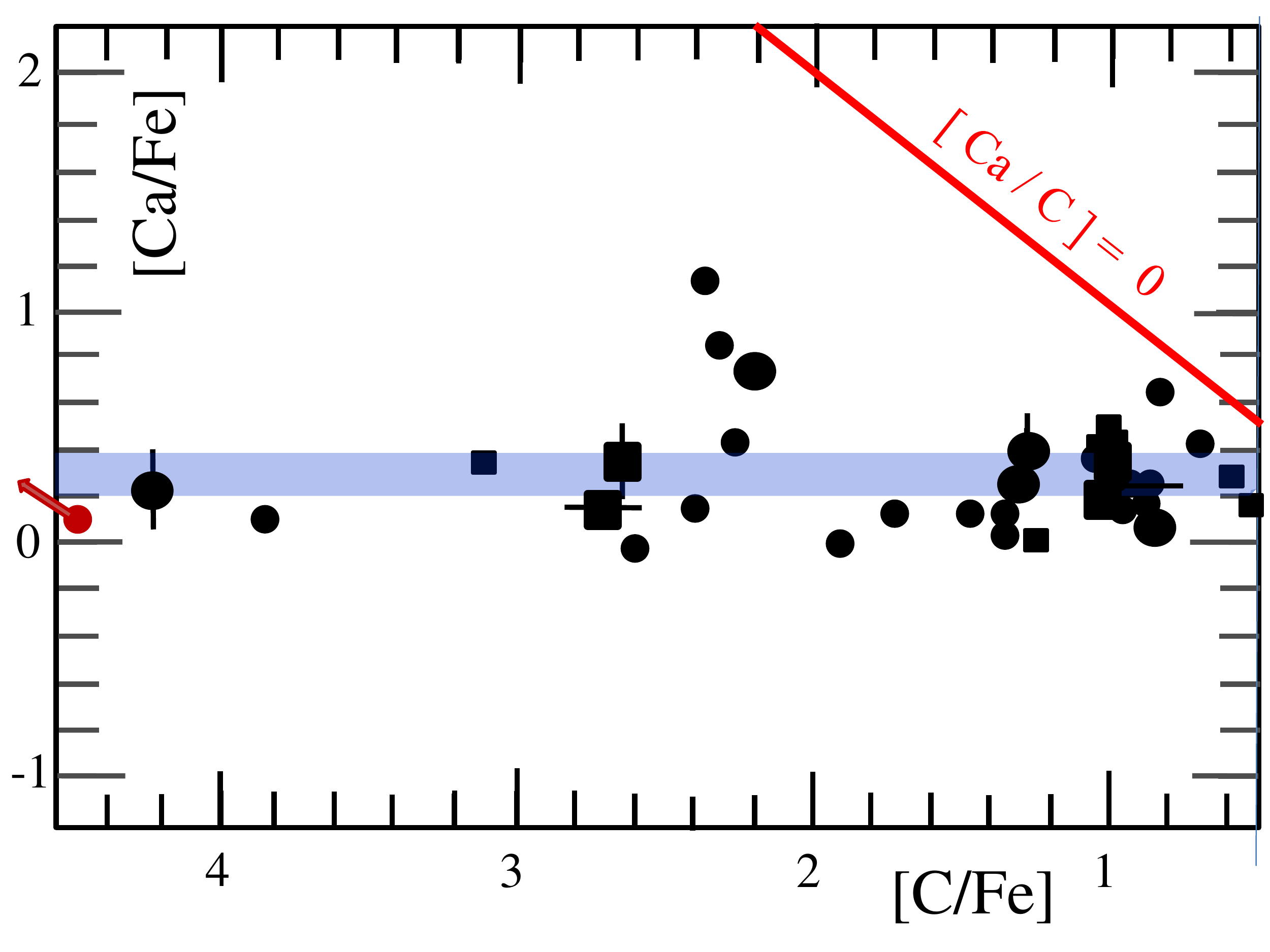}
\caption{Abundance ratios [Ca/Fe] vs. [C/Fe] of CEMP--no stars.  The same codes for the observations and models as in
 Fig. \ref{C12C13} are adopted; no data on Ca are provided by \citet{Masseron2010}.  A thick horizontal blue line indicates the approximate 
average of the [Ca/Fe] ratios. The location corresponding to a solar ratio [Ca/C]= 0 is shown with a red line.
  The star SMSS 0313-6708 \citep{Keller2014} is represented by a brown point and is a lower limit on both coordinates.}
\label{CAFECFE}
\end{center}
\end{figure}

\section{Relation of CNO to $\alpha$-elements}   \label{ca}

 CEMP--no stars bear the traces of  $\alpha$--elements, such as Si, S, and  Ca, which are formed
in the onion-skin layers of the pre-supernovae. The strongly enhanced
CNO elements, which make CEMP--no stars so particular, necessarily imply some other major effect(s), as mentioned above.
In this section, we perform  additional tests showing the huge difference between  the properties of the $\alpha$--elements, such as calcium and silicon,  and those of the  
C, N,  and O elements. The interest of the abundances of  elements like Si and Ca, produced deeper in stars than are the lighter elements,
has already been emphasized by \citet{Norris4} in relation to the possibility of distinguishing between spinstar and mixing and fallback models to account for the properties of CEMP--no stars. These authors also emphasize the need "to have more accurate abundances in a larger sample of C--rich stars for comparison with more detailed model predictions for the two classes of models". 

  Figure \ref{CAFECFE} shows the [Ca/Fe] ratios vs. the [C/Fe] ratios for CEMP--no stars. In this plot the stars with low [Fe/H] are on the
left, since [C/Fe] is higher for lower [Fe/H] ratios. We see no correlation between   calcium and  carbon abundances. 
The  scatter  around an average of [Ca/Fe] = 0.2 to 0.3 is small ($\approx 0.2$ dex, {\emph{i.e.,}}  about the same size as the observational errors).
This flat curve  means that the enrichments in Ca were the same as those in Fe from the core of the source stars,
while   carbon was, incredibly, produced  much more than Fe and Ca at very low [Fe/H] values.  The average [Ca/Fe] ratio around 0.2 is the same as the one observed in normal halo stars.               
While the [Ca/Fe] ratios of CEMP--no stars are within a range of about 1 dex  (most  stars lying within a range of 0.5 dex), the  [C/Fe]  values cover a range of 4 dex! Inspection of the figures representing the standard classical chemical yields of supernovae \citep{Maeder1992,Hirschi2005,Hirschi2007} clearly shows that classic supernova models 
cannot  produce such  large changes of [C/Fe] while producing such limited changes  of  [Ca/Fe].

\begin{figure}[t]
\begin{center}
\includegraphics[width=9.1cm, height=7.0cm]{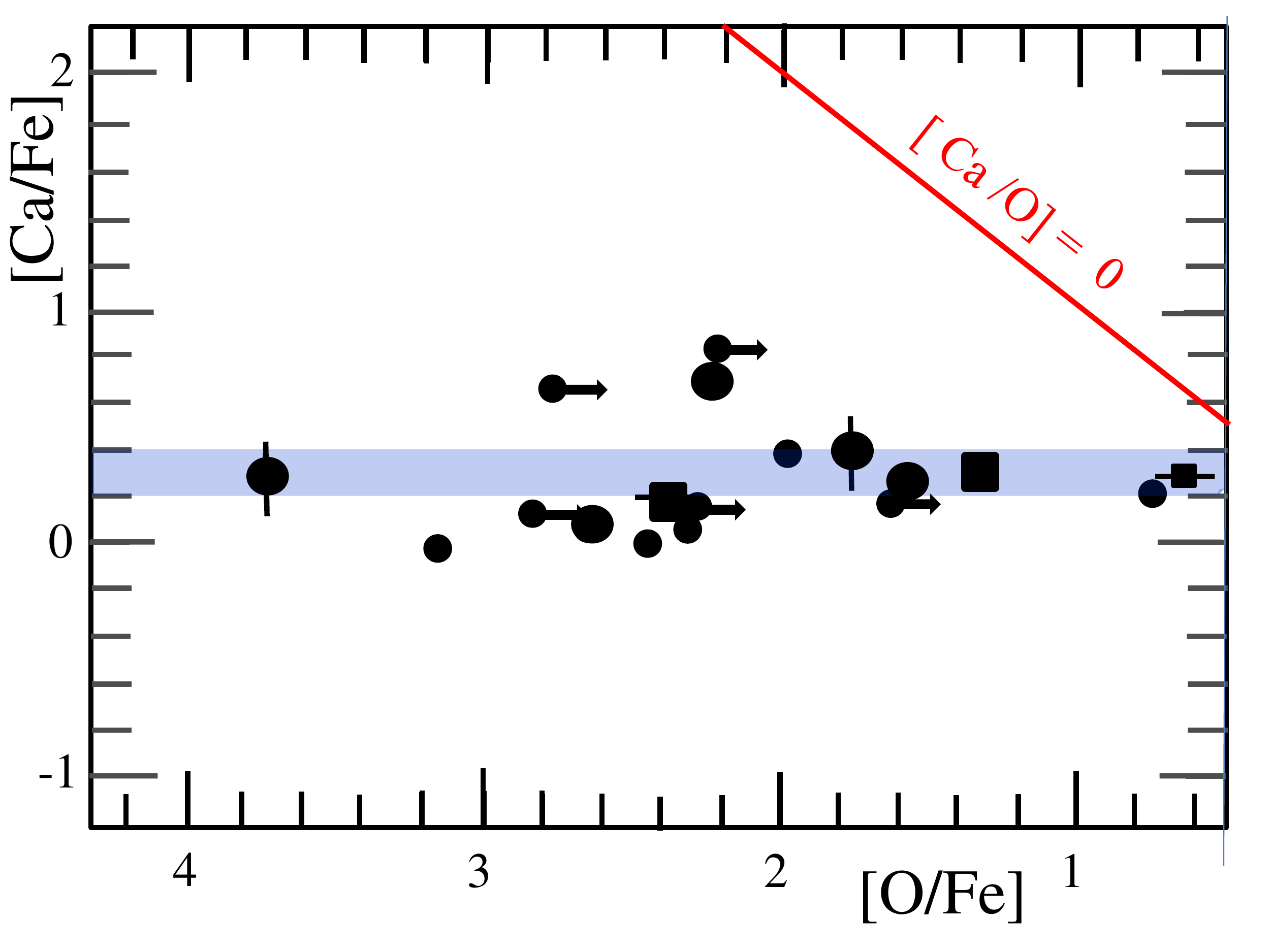}
\caption{Abundance ratios [Ca/Fe] vs. [O/Fe] of CEMP--no stars.  The stars are not all the same ones as in Fig. \ref{CAFECFE}, because not all stars that have C--data necessarily have O--data.
 A thick horizontal blue line indicates the approximate 
average of the [Ca/Fe] ratios. The location corresponding to a solar ratio [Ca/O]= 0 is shown with a red line.}
\label{CAFEOFE}
\end{center}
\end{figure}

Figure \ref{CAFEOFE} shows the  [Ca/Fe]  vs. the [O/Fe] ratios. The same  absence of correlation  emerges as before. This was not evident since in pre--supernova models the O-rich layer lies  deeper  inside than the C rich layer and is  the thickest one, making oxygen the most abundant of the $\alpha$--elements. Thus, in the framework of classical supernova models, it is surprising that the behavior of oxygen has absolutely nothing to do with calcium's behavior.  In this connection, we may point out that supernovae models for Population III
stars, as well as those  for very low $Z$ stars,  show  a positive relation between the Ca and O productions (see Fig. 5 by \citet{Nomotoaraa2013}.
In addition, as in the case of C, the ratio [O/Fe]  varies by  4 dex, while the range of [Ca/Fe] values is very limited. 


Figure \ref{SIC} shows the same kind of relation between [Si/Fe] and [C/Fe]. In pre-supernovae models, the silicon layer lies above the calcium layer and is close to the oxygen one. Again, there is no correlation between [Si/Fe] and [C/Fe], and the ranges covered by [Si/Fe] and [C/Fe] are very different, as in the two previous figures. This further confirms the completely different origin of the CNO  and $\alpha$--elements. 
As a matter of fact, the [Ca/Fe] and [Si/Fe] ratios of CEMP-no stars are fairly similar to the ones found in normal halo stars (see Fig. 4 in \citet{Chiappini2013}.

\citet{Norris4}  noticed that some relative  enhancements of Si and Ca exist in some
CEMP--no stars. Indeed, from Table 1 we see that  some stars that have a high [Si/Fe] abundance ratio also have a hig{h [Ca/Fe] content, for example HE 1012-150 and Segue 1-7. The data of Table 1 also show that there are   stars with a high [Si/Fe]  and  a relatively low [Ca/Fe] ratio, such as CS 29498-043 and  HE 2139-5432.  We notice that possible differences between the behaviors of Ca and Si (or between other $\alpha$--elements, as well as  r--elements) are particularly interesting with respect to  the parameters concerning the physics of supernova models, such as mixing and fallback models. A parameter like the mass cutoff may strongly influence the abundance ratios of elements produced deep inside. A study of these effects is beyond the scope of this paper.

The results of Figs. \ref{CAFECFE}, \ref{CAFEOFE}, and \ref{SIC}  clearly confirm the
huge differences in the enrichments of CEMP--no stars  in CNO--elements and  $\alpha$--elements (other than oxygen)
and the relative lack of the $\alpha$--elements at the very low [Fe/H].
These elements have necessarily been produced in the pre--supernova stages. Thus, where are they? We think there
    are essentially two hypotheses to explain this situation: either the stellar remnants retain  the $\alpha$--rich onion skin layers, 
as assumed in the mixing and fallback models,
 or these $\alpha$ layers,  ejected at high velocities, partially escape from the outer galactic halo where CEMP--no stars went on to form. 
Whether the presence of small amounts of heavy elements and the differences between these heavy elements  favor
one of the two possible proposed explanations remains to be investigated.

\begin{figure}[t!]
\begin{center}
\includegraphics[width=9.1cm, height=7.0cm]{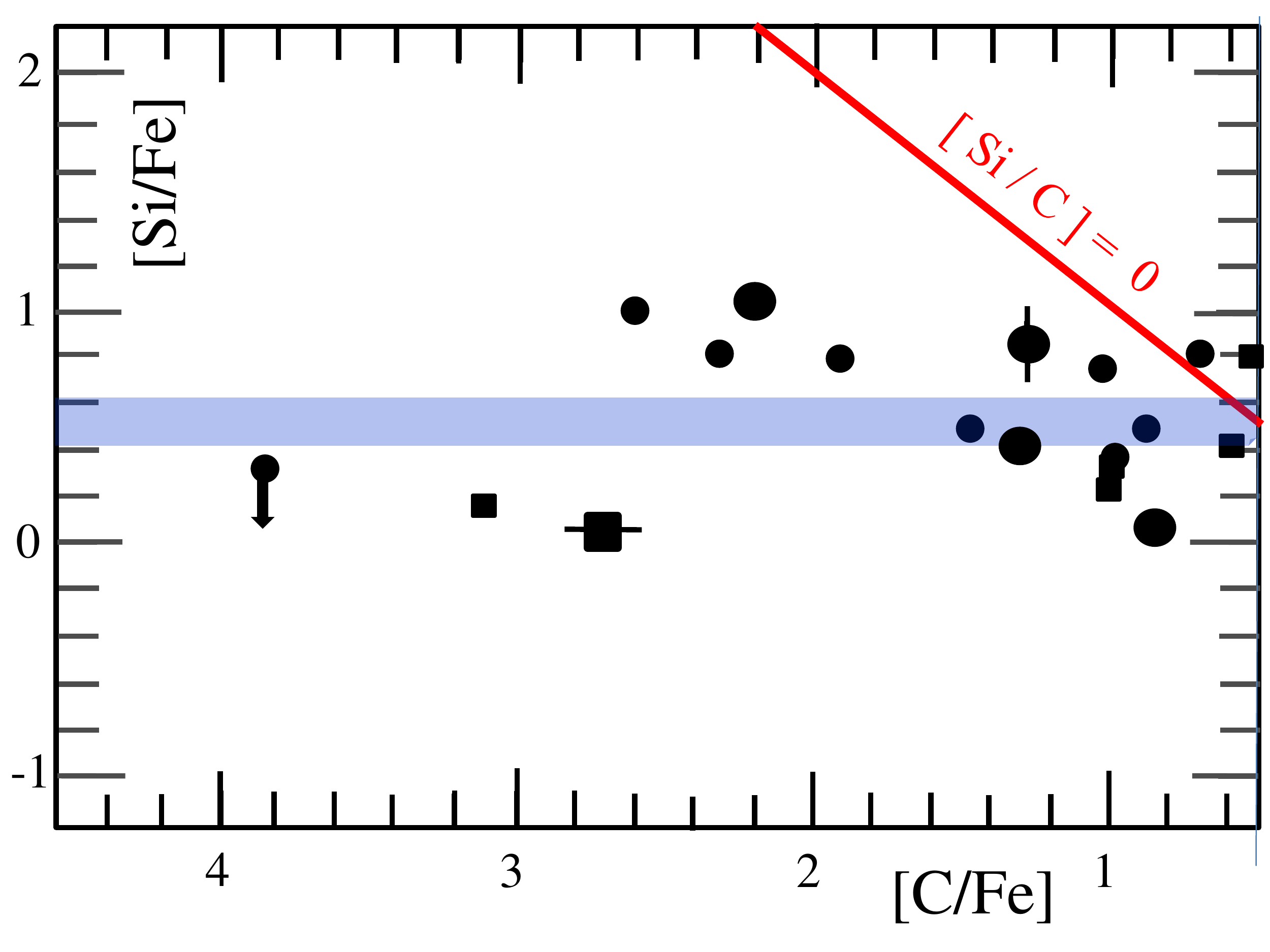}
\caption{Abundance ratios [Si/Fe] vs. [C/Fe] of CEMP--no stars.  The stars are not all the same ones as in Fig. \ref{CAFECFE}, because not all stars that have Ca data necessarily have Si data.
  A thick horizontal blue  line indicates the approximate 
average of the [Si/Fe] ratios. The location corresponding to a solar ratio [Si/C]= 0 is shown with a red line.}
\label{SIC}
\end{center}
\end{figure}

\section{The elements of the Ne--Na and Mg--Al cycles}  \label{nenamgal}

These two cycles,  which accompany the  CNO cycles, are not significant for energy production, but they do influence some abundances and isotopic ratios \citep{maederlivre09}. If the anomalous abundances in CEMP--no stars are products of He--burning (mainly C, O, $^{22}$Ne, etc.) partially gone through CNO H--burning, there should also be some effects visible in the elements participating in the Ne--Na and Mg--Al cycles,
which proceed by successive $(p,\gamma)$ reactions and  $e^{+}$ emissions.  Thus, we may expect some specific relations between the
elements  involved in these cycles, especially more since the light elements Na, Mg, and Al are enhanced relative to Fe in about the half of the CEMP--no stars, as  shown by \citet{Norris4}. Also there should be 
some broad distributions of the  concerned chemical abundances, rather than  narrow  ones as seen above for Ca and Si.
Several nuclear rates involved in the two cycles are still highly uncertain, so that detailed quantitative predictions are difficult. As shown by \citet{Decressin2007}, some rates have to be "pushed" to their possible limits and even increased by a factor of 1000 in order to account for some anticorrelations of abundances observed in globular clusters.

\subsection{The Ne--Na cycle}    \label{nena}

The  main effect of the Ne--Na cycle is to transform  the various Ne isotopes ($^{20}$Ne, $^{21}$Ne, $^{22}$Ne) into $^{23}$Na  at relatively high temperatures above $4 \cdot 10^7$ K.  Excesses of sodium have been found in massive supergiants by 
 \citet{Boyarchuk1988} and further confirmed by many works. The Ne--Na cycle is also responsible of the "Na--O anticorrelation" observed for  the stars in some globular clusters  \citep{Gratton2001}. Many stellar sites have been proposed for this effect, most noticeably massive stars \citep{Decressin2007}. The anticorrelation is likely the result  of
 H-burning,  during which Ne is turned by $(p,\gamma)$ reactions to Na by the Ne--Na cycle. At the same time, 
 O is also transformed into N by the ON loops of the CNO cycle, thus creating the observed anticorrelation. In a  cluster, the importance of mixing is different for each star so that different amounts of Na and CNO elements become visible at the surface, and
this variety in the transport process   allows one  to see some relations between different elements. This may be the effect at the origin of the Na-O anticorrelation, observed   for stars within a given globular cluster, although a variety of interpretations have been proposed (see
\citet{Decressin2007} for further references). \\

The problem of the CEMP--no stars is different. The CEMP--no stars cover a very wide range of [Fe/H] values, from [Fe/H] = -7.1 to -2.5. The abundances of the  
CNO--elements  may vary by a factor of  $10^4$ or more. These variations largely dominate the limited amplitudes  of the observed Na--O and Mg--Al anticorrelations,  which amount to less than 1 dex (and even 0.4 dex for Mg and Al), according to  \citet{Gratton2001}.  Thus, the possible anticorrelations are largely overwhelmed  by the global variations in the CNO  as a function of [Fe/H], which may be up to 1000 times larger   for CEMP--no stars. The Ne--Na chain reactions, active in the H--burning shell,  only contribute to increasing the scatter around the general trends,
scatter that is not negligible as we will see.
\\

\begin{figure}[t]
\begin{center}
\includegraphics[width=9.1cm, height=7.0cm]{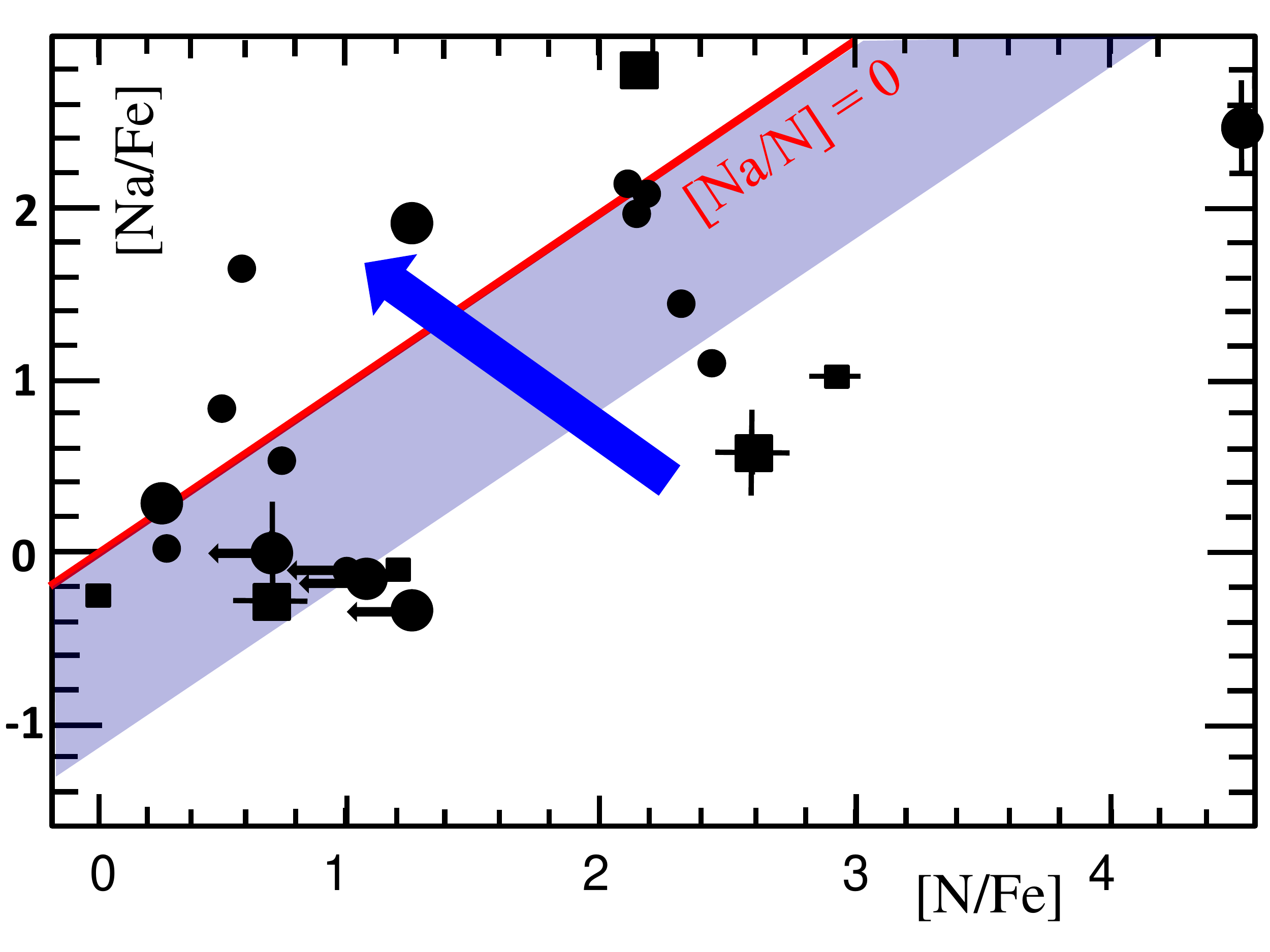}
\caption{abundance ratios [Na/Fe] vs. [N/Fe] of CEMP--no stars.  The dots represent the star by \citet{Norris4}, the squares by \citet{Allen2012}. The small vertical bars on the points indicate  the  main sequence or subgiant  Li-poor stars.  
The small horizontal bars indicate the low--s stars  (also noted b-).
The light thick blue band  represents the  main trend of the observed points.
The big blue arrow describes the effect of nuclear burning of  $^{14}N \, \rightarrow  \, ^{22}Ne \, \rightarrow  \, ^{23}Na$, which contributes to the scatter of the figure. Small arrows indicate upper limits.}
\label{NAN}  
\end{center}
\end{figure}

For examining the Ne--Na cycle in the absence of Ne--data, we  try to use  the N data, because of the strong relations between nitrogen and neon. First, at the beginning of   He--burning in the core,  $^{14}$N is transformed into $^{22}$Ne. Then, the neon that diffuses  to the H--burning shell is transformed by the Ne--Na cycle to $^{23}$Na. 
 As the He--burning further 
proceeds, some amounts of $^{20}$Ne are  created, which by diffusion in the H--burning region may also participate in the Ne--Na cycle (see for example Fig. 27.17 by \citet{maederlivre09}.

Figure \ref{NAN} shows the relation between the [Na/Fe] and [N/Fe] ratios. We notice the following properties:
\begin{itemize}
\item  There is some  correlation between the abundances of Na and N. The scatter is the same for MS stars and giants, which suggests
that self--enrichment does not significantly influence this plot. The scatter is relatively large, which is not surprising  since the nuclear-burning
chain  $^{14}N \, \rightarrow  \, ^{22}Ne \, \rightarrow  \, ^{23}Na$ acts almost perpendicularly to the main  trend of the data, as shown in 
Fig. \ref{NAN}.
\item  Despite the scatter, we see that the global slope of the main trend is not far from 1. This means that the ratio  Na/N  stays almost constant through the range of CEMP--no stars, a fact
 consistent with the operation of the Ne--Na cycle.
\item The main trend  results from the very large variation in the [N/Fe] among the CEMP--no stars: stars that have lots of N also produce lots of Na. This effect, which spans several dex, overwhelms the
fact that at a given metallicity, the partial  production of $^{23}$Na implies some  $^{14}$N--destruction. This last effect
only contributes to increasing the scatter  of the general relation, as illustrated in Fig. \ref{NAN}. 
\item The abundances of Na vary by orders of magnitude, as is the case for C, N, and O elements. This  is    much more than the variations  in Ca and Si, typically produced in the onion-skin layers of supernovae. Thus, the amplitude of the variations suggests that at least a large portion of the
Na in CEMP--no stars has an origin related to the H--burning cycles. 
\end{itemize}

\subsection{The Mg--Al cycle}

The main effect of the Mg--Al cycle is to transform, also by  $(p,\gamma)$ reactions and $e^{+}$ emissions, the various $^{24,25,26}$Mg isotopes
into $^{27}$Al. (This cycle also produces  the long--lived radioactive $^{26}$Al,  responsible for the  galactic $\gamma$--emission.) The cycle operates in H--burning regions typically at temperatures above $5 \cdot 10^7$ K \citep{Decressin2007}. An anticorrelation  Mg--Al has  been
observed  in globular clusters \citep{Gratton2001}. It has generally been interpreted in terms of the Mg--Al cycle: as Mg is destroyed by  $(p,\gamma)$ reactions, new Al is formed. Various kinds of stars have been considered as a potential   site of these reactions, the case of massive stars with internal mixing has  been studied by  \citet{Decressin2007}.

Here,  we have data for the relevant elements of the Mg--Al cycle.   Figure \ref{MGFEALFE} illustrates the 
[Al/Fe] vs. [Mg/Fe] relation for CEMP--no stars. The blue arrow indicates the direction
of the burning of Mg to Al, and this effect also contributes to the scatter around the general trend.   We note the following facts:
\begin{itemize}
\item Both Mg and Al increase simultaneously (for lower [Fe/H]) with  a slope of about 1.0.  This means that the ratio of Mg to Al  stays  about constant  throughout the range of CEMP--no stars, a fact that is consistent with the operation of the Mg--Al cycle for these stars.
There are two other arguments  that support the view that Al is mainly due to the destruction of Mg
produced in the helium-burning core.

\begin{figure}[t]
\begin{center}
\includegraphics[width=9.2cm, height=7.0cm]{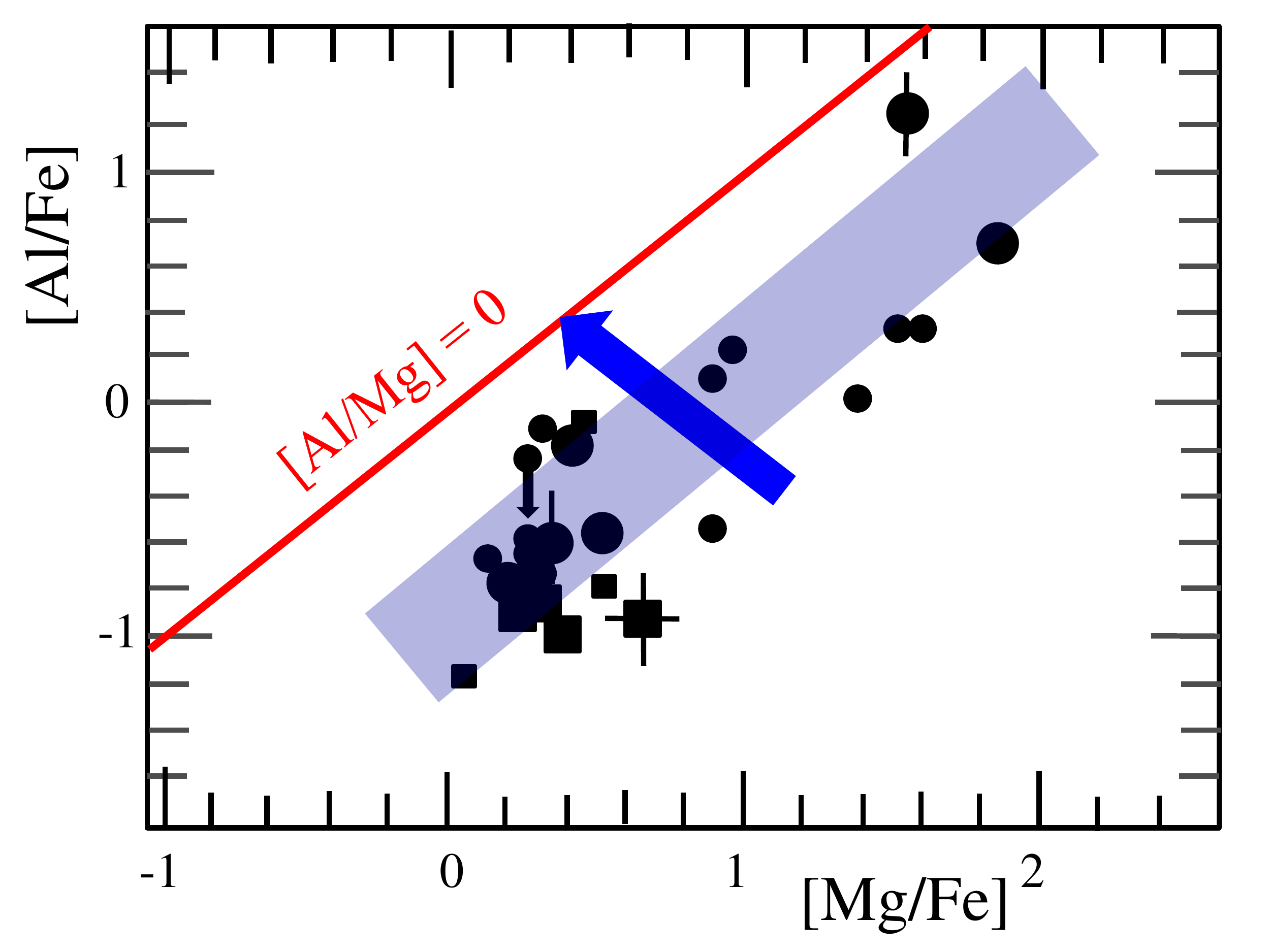}
\caption{Abundance ratios [Al/Fe] vs. [Mg/Fe] of CEMP--no stars.  
 The wide light blue  band describes the main trend. The  thick blue arrow
illustrates the direction of the burning by the Mg--Al cycle. Same remarks as for Fig. \ref{NAN}.} 
\label{MGFEALFE}
\end{center}
\end{figure}

\item First, models show that the high Mg abundance needed before the Mg-Al cycle comes into play
may be
 created in low $Z$ stars during the final part of the He-burning  phase \citep{Meynet2006}. A fraction of this element may be  conveyed to the stellar surface, where it may escape in the winds.  This is illustrated well  by
models of rotating massive stars with $Z=10^{-5}$ by \citet{Meynet2006}, which  show
that the winds present  enrichments in   Mg (summing over the 3 isotopes) by a factor 24 with respect to the initial Mg content, as shown in Table 4 of the mentioned reference.
\item Second, the synthesis of Mg near the end of the  He--burning phase of low $Z$ stars does not exclude some production of this element 
by supernovae. However, the amplitudes of the variations in
Mg and Al do not plead so much for this hypothesis as a dominant process for these elements. 
 In  Fig. \ref{MGFEALFE}, both elements  Mg and Al vary by two orders of magnitude. This is more than the
cases of Si and Ca, which are ejected at the time of the supernova explosion.

\item Thus, the relation shown in Fig. \ref{MGFEALFE}, together with amplitudes of the variations in the abundances of Mg and Al,
supports a scenario in which the largest part of these two elements comes from the migration of core
He-burning products into the H-burning shell.
\end{itemize}

 The "mother-daughter" relations with slopes of about 1.0 for N-Na (more a relation between  grand mother and grand daughter) and Mg--Al are in agreement with partial H-burning of some products (Ne, Mg) of He-burning by the CNO  and related cycles.
That we do not observe the anticorrelations present in globular clusters is not at all contradictory, because the CEMP--no star composition is not the result of H--processing alone, but also the result of material
produced by He--burning  (in very variable amounts) and partially further transformed by H--burning. 
The extremely  broad range 
of the  CNO elements (and Na, Mg, and Al)  present in CEMP--no stars largely overwhelms  the effect of the H--burning cycles producing the
anticorrelations. 

In summary, the present analysis strongly suggests that  elements such as Na, Mg, and Al, which are present in CEMP--no stars,
behave like the CNO elements. These elements   are thus produced by the Ne--Na and Mg--Al cycles operating in H--burning regions of the source stars.

\section{An attempt to organize the CEMP--no stars}  \label{organize}

\begin{table*}[t!]  
\vspace*{2mm}
 \caption{Possible steps in the physical processes of partial mixing and nuclear H-- and He--burning in massive very low $Z$ stars. 
He should be enriched  in all cases, except for step number 0. See text for more details. } \label{steps}
\begin{center}   \footnotesize 
\begin{tabular}{cccc} 
 Steps & Acting physical processes & Main nuclear products & Stars and properties   \\
&   &  & \\
\hline
&   &  & \\
&   &  & \\
  0  & Absence of mixing &   Initial composition   & Low $Z$ stars without C excess \\
0+ &Core H-burning & $^{13}$C, $^{14}$N & Very low CNO, but  N>C,O  \\
  &  &   &  \\
    1  &  Core He-burning    & $^{12}$C, $^{16}$O, no N &  Visible only in WC stars\\   
1+ & Advanced He-burning  & $^{12}$C $\searrow$,  $^{16}$O, low  $^{20}$Ne & Visible only in WO stars \\
&   &  & \\
 2  &  Partial mixing   $^{12}$C,  $^{16}$O  in H-burning shell &$^{12}$C, $^{13}$C, $^{14}$N,$^{16}$O & CEMP-no  with [C/N]>0, [O/N]>0\\   
2+ &  Idem with  advanced H-burning  & Idem  & CEMP-no with [C/N]<0, [O/N]<0\\
2Na & Idem from 1+ with the Ne-Na cycle   & Idem, $^{20,21,22}$Ne, $^{23}$Na & CEMP-no, [C/N]>0, [O/N]>0,Na\\   
2+Na & Idem 2Na, advanced H-burning   & Idem, $^{20,21,22}$Ne, $^{23}$Na & CEMP-no, [C/N]<0, [O/N]<0, Na\\
&  & & \\
3  & Partial mixing prod. (2, 2+)  in He--burn. zone   &  $^{18}$O, $^{22}$Ne &   CEMP-no, N $\searrow$, O-strong,$^{22}$Ne \\ 
3+ & Idem with advanced $\alpha$ captures  & O,$^{20,22}$Ne,$^{25,26}$Mg,Sr,Y & CEMP-no,idem,Mg,s-elem.(1$^{st}$peak)\\
3++ & Idem with more $^{22}$Ne ($\alpha$, n)$^{25}$Mg captures & Idem, Ba weak &Low s-star, Mg,s-el. (1$^{st}$,2$^{nd}$ peak)\\
     &   {\emph{The same  above 3 cases from (2Na,2Na+)}} &   & {\emph{Same properties as  in above cases}}  \\
     &{\emph{$\longrightarrow$ cases 3Na, 3+Na, 3++Na}} &   & {\emph{with Na present }} \\
  &   &    &\\ 
4 & Part. mix. 3 in H-burn., CNO,Ne-Na  & CNO, Ne,Na, & CEMP-no, N$\nearrow$, Ne, Na,\\
4+ & Part. mix. 3+ in H-burn., CNO,Ne-Na,Mg-Al   & CNO, Ne,Na,Mg,Al & CEMP-no,Na,Mg,Al,s-el.(1$^{st}$peak)\\
4++& Partial mixing of 3++, idem  & Idem            & Low s, Na,Mg,Al,s-el.(1$^{st}$,2$^{nd}$peak)\\
& & & \\
&   &  & \\
   \hline 
\normalsize
\end{tabular}
\end{center}
\vspace*{0mm}
\end{table*}

In natural sciences, a first step toward a better  understanding often goes through classification of the objects under investigation. Since the extraordinary peculiarities of CEMP--no stars  result from the partial mixing of  products of He--burning  into the H--burning shell, before escaping from the source stars, we may use this property to build a classification scheme by considering successive steps in the process of mixing. We note that the timescale of mixing 
is relatively short for  rotating stars \citep{maederlivre09} so that a fraction
of the newly synthesized elements can make a few back-and-forth motions between the He-- and H--burning regions. This may result in further nuclear chains operating in the layers where some of the new  
elements enter. Mixing is always partial, affecting only a part of the new elements synthesized. Some H--burning reactions, such as the  Ne--Na and Mg--Al cycles, cannot operate  as long as some  isotopes have not been created by the succession of back-and-forth matter exchanges  between the H--
 and He--burning layers.

\subsection{A possible classification scheme}
Table \ref{steps} is an attempt to establish a classification on the basis of simple considerations based on the  reactions involved in 
H-- and He--burning regions and mixing between them. From the initial, very low $Z$ composition ($Z \geq 10^{-9}$), if mixing brings   
some  products of H--burning  to the surface, the star keeps its very low CNO content, but with N enrichments (step  0+). These are just EMP--stars. Steps 1 and 1+  consist of stars where pure products of He--burning become visible at the stellar surface. In  case 1+, the $(\alpha,\gamma)$ reactions lead to the production of some $^{20}$Ne, in addition to small amounts of $^{22}$Ne resulting from $^{14}$N.
These 1 and 1+ objects are stars like WC and WO stars. It is still uncertain whether such objects exist at extremely low metallicities.

Steps 2 and  2+ consist of stars where some quantities of C and O from the He--burning core (case 1)  have been partially
mixed into the H--burning shell, where partial burning is forming some $^{14}$N. Then large amounts of CNO become visible at the stellar surface 
as a result of mixing and/or mass loss. We distinguish  Case 2 where mixing is very  mild (or H--burning proceeds weakly), which leads to
  [C/N] > 0  and [O/N] > 0 and Case 2+ where the H--burning of the mixed matter is more complete, leading to ratios [C/N] < 0 and [O/N] < 0, as well as to low $^{13}$C/$^{12}$C ratios. Steps 2Na and 2+Na are just the same as the previous ones, but starting from Case 1+, where
some Ne has also been produced. If the  temperature is high enough, this allows the Ne--Na cycle to operate  and to produce some $^{23}$Na, which may be  observable.

In Steps 3 and after, we consider that some products of Steps 2 and 2+ are again mixed from the H--burning into the He--burning region, leading to new element synthesis, in particular from the reactions,
\begin{eqnarray}
^{14}\mathrm{N}(\alpha,\gamma)^{18}\mathrm{F}(\,,e^+ \nu_{\mathrm{e}})^{18}\mathrm{O}(\alpha,\gamma)^{22}\mathrm{Ne} \longrightarrow \; \;
(\alpha, n)^{25}\mathrm{Mg}  \quad \quad  \\[1mm]
 \searrow   \; \;(\alpha, \gamma)^{26}\mathrm{Mg} \, .  \quad \;  \;
\label{ne22}
\end{eqnarray}
\noindent
 In the first case (Step 3),  we consider that reaction (10)  leads to a small amount of the relatively rare isotope $^{18}$O and then goes up  to $^{22}$Ne, without reaching a significant  formation of $^{25,26}$Mg isotopes, the formation of which requires  higher temperatures. These more advanced stages 
of $(\alpha,\gamma)$ and $(\alpha, n)$ captures  occur in Step 3+ with the complete reactions (10) and (11) operating. At the same time,
 there is  some formation  of  $^{20}$Ne from the reaction
\begin{eqnarray}
^{16}{\mathrm{O}}(\alpha,\gamma)^{20}{\mathrm{Ne}} \, . 
\end{eqnarray}
The capture of neutrons coming from reaction (11) leads to the formation of some s--elements  from the first peak, like strontium Sr and yttrium Y,
as shown by \citet{Fris2012}. In Step 3++, the captures of neutrons from reaction (10) goes further in the production of s--elements with, in addition to the s--elements of the first peak, a significant production of s--elements from the second peak, such as barium Ba and lanthanum  La. 
The star is
likely no longer considered  as a CEMP--no star, according to the current classification criteria, but rather as a low s--star or a "b-" \citep{Masseron2010,Allen2012,Norris4}. 

When the products of partial mixing into the He--burning zone are those coming from Steps 2Na and 2+Na, the same reactions as the previous ones 
may occur, leading to the same nucleosynthetic products, simply with Na present  in addition to those already mentioned. This leads to Steps
3Na, 3+Na and 3++Na as indicated in Table 2.

In Step 4, some elements of Step 3 are again   mixed into the H--burning zone, and this may  allow the Ne--Na cycle to operate and produce
some Na. Simultaneously some N is also synthesized  from the C and O present in Step 3, and its abundance may be high again.
In Steps 4+ and 4++, some elements of the corresponding Stages  3+ and 3++ (with or without Na) are brought
into the H--burning region, where now the cycles Ne--Na and Mg--Al are both operating. This then leads to the presence of
all the isotopes of elements  Ne, Na, Mg, and Al occurring  in  these cycles, in addition to those created in the previous phases including the 
s--elements. 

The main conclusion of what precedes  is  the extraordinary variety of compositions that may result from partial mixing between the H-- and He--burning phases.
In Table 2, there are 17 different cases that are theoretically possible. We could even have considered more of them, for example, by introducing 
further distinctions based on [C/N] and [O/N] ratios that are greater or less than 0, as we did in Steps 2 where this seems appropriate.
It is  possible that we do not find stars corresponding to some of the cases we have  considered.   Future works may hopefully try to 
analyze the observed CEMP--no and low s--stars in terms of the proposed scheme. Even some stars with a significant [Ba/Fe] ratio might 
belong to the genetic group of the descendants of spinstars.  More objects with accurate data  are certainly 
necessary for this undertaking, which may also lead to improvements in the proposed classification scheme.

\subsection{The stars with [Sr/Fe]  > 0} \label{sr}

In principle, CEMP--no stars contain  no or few s--elements. The ratios [Sr/Fe]  and [Ba/Fe] are generally negative \citep{Norris4}. Sr, like Rb, Y, Zr, 
Nb, etc., are s--elements of the first peak (corresponding to "a magic number" of neutrons N=50). Ba, like Cs, La, Ce, etc., are  s--elements of the second peak (corresponding to a magic number N=82). There are four stars in Table 1 with positive [Sr/Fe] ratios, which  we now discuss briefly.

 The star HE 1327-2326  has  [Sr/Fe]= 1.04, the highest ratio in the sample, and  for barium an upper limit [Ba/Fe] < 1.46 \citep{Norris4}, which is 
formally too high for a CEMP--no star, as already mentioned.
 We note that this star  has a  high [Mg/Fe] ratio of 1.55, consistent with the operation of the 
$\alpha$--capture  by $^{22}$Ne and neutron emission producing the  synthesis of the s--elements. 
 In this context, we recall that 
the models of massive rotating stars by \citet{Fris2012}  show that, although the s--elements produced  mainly belong to the first peak,
 there are cases, especially for high rotation velocities, where a substantial number of elements  of the second peak  are formed. Thus, HE 1327-2326 is probably a member of the same genetic family
as the formal CEMP--no stars, simply with much  more complete mixing and nuclear burning than the average. This view is  well supported by the fact that the other significant ratios are  very high, with  [Na/Fe]=2.48, [Mg/Fe]=1.55 and [Al/Fe]= 1.23, which    indicates that relatively large quantities of  the elements  of the Ne--Na and 
Mg--Al cycles are reaching the stellar  surface. In terms of our classification of Table 2, we would consider it as a 4++ star, which is the extreme case. 

The three stars other with positive [Sr/Fe] values are CS 22949-037, BS 16929-005, and 53327-2044-515, with [Sr/Fe]= 0.55, 0.54, 0.81,
and [Ba/Fe]= -0.52, -0.41, < 0.15, respectively \citep{Norris4}. Thus, these stars essentially  have s--elements of the first 
peak and  little from the second peak (in agreement with the formal definition of a CEMP--no star). Also, the lower Mg content of these
stars (especially the last two) is consistent with fewer s--elements. The small abundances of aluminum indicate a relative
absence of the Mg--Al cycle for  the three stars, while the products of the Ne-Na cycle are visible in CS 22949-037.

 The  beautiful variety of the compositions of CEMP--no stars is likely to result from  the intensity of the various nuclear reaction 
chains and cycles, as well as from the importance of mixing and mass loss in the source stars. These different effects   probably depend on
the  different values of the  stellar parameters, such as the initial masses (above
about 7 M$_{\odot}$),  the different [Fe/H] ratios,    and rotational velocities.

\section{Conclusions}    \label{concl}
  The extraordinary properties of the CEMP--no stars are a gift of Nature to help us to try to understand the first  generations of stars in the Universe.  We suggest  the following tentative conclusions and lines
of further exploration.

-- Chemical abundances support the view that CEMP--no stars exhibit products of He--burning that has partially gone through CNO processing. A portion of the elements may  have experienced successive back-and-forth motions between the He-- and H--burning zones in the source stars.

-- The comparison of the  $^{12}$C/$^{13}$C vs. [C/N] ratios  of models and observations   showed that the mass transfer in binaries with an AGB star is an unlikely source  for the CEMP--no stars. This further confirms partial mixing, because   full mixing
would have  destroyed  the $^{13}$C  isotopes.

-- Models of massive rotating stars with partial mixing  and nuclear processing, and with strong  mass loss in the He--burning and later phases, agree relatively well with the observed compositions.  In this context, we note that  the solar ratios  should not be used to define the initial abundances ratios of the CNO  and related elements, because this creates  some artificial differences between models and observations of CEMP--no stars. Future models that take this remark into account may have more discriminating power. 

-- As a marginal remark, we point out that  four stars in the sample have $\log g > 4.60$, which is surprising for such old objects.
 This may support the hypothesis of a relatively high density for these objects, owing to high He--content.

-- We confirm the finding by \citet{Masseron2010} of a  linear relation between  the [(C+N+O)/H]  and [Fe/H] ratios for CEMP--no stars down
to  [Fe/H] $\approx$ -3.5 or  - 4.0. This means that the  [(C+N+O)/Fe] ratios are constant,  indicating a primary behavior of the CNO elements in the considered range of [Fe/H].

-- Below  [Fe/H]= - 4.0, the ratios  [(C+N+O)/H] appear to no longer vary the same way, since they are either constant or show only a slight decrease with [Fe/H], so that  the  [(C+N+O)/Fe] ratios become very high, reaching about 
4 dex. These values may just correspond to individual nucleosynthetic properties of the source  stars, because in view of the timescales, there is no effective   average galactic chemical evolution for such low [Fe/H].

-- The $\alpha$--elements with atomic mass number > 24, such as Si and Ca,  present properties that are completely different from those of  the CNO, Ne--Na, and Mg--Al elements: the ratios  [$\alpha$-- elements/Fe] stay constant with a small scatter,
while CNO and related elements show strong increases of the  [CNO,.../Fe] ratios for  lower [Fe/H]. These behaviors 
suggest that the  heavy elements of the onion-skin layers of the presupernovae  contributed less than the stellar winds from partially mixed stars in the earliest phases of the chemical enrichment of the Galaxy. Some possible reasons for this effect were proposed.

-- The elements Na, Mg, and Al show a range of variations, which is  much greater than for the $\alpha$--elements and is more similar to that of the CNO elements. Nevertheless,  [Na/N] and [Al/Mg] stay about constant for the low [Fe/H] ratios. This supports the view
that the Ne--Na and Mg--Al  cycles  were efficiently operating in the source stars. 

-- The CNO, Ne--Na, and Mg--Al elements of CEMP--no stars present a wide variety of abundances. Continuous partial mixing leading some
matter to experience several consecutive phases of He-- and H-- burning may explain this variety. We propose a  classification
of CEMP--no stars based on the properties of these successive phases.

-- The succession of the  phases of He-- and H--burning may lead to the formation of s--elements, mainly of    
the first peak with Sr and Y. In case of sustained neutron emissions from the reaction $^{22}$Ne ($\alpha$,n)$^{25}$Ne,
some elements of the second s--peak may also be produced.  In the sample of Table 1, there are four such stars with [Sr/Fe] > 0,
and one  of them (HE 1327-2326) may also show significant Ba--enrichment. Even, if it is not formally cataloged as a CEMP--no star according to the
current criteria, it  belongs to the same family, showing all its characteristics \citep{Norris4}.

There is a strong need  for accurate observations of more stars, as well as for more complete grids of detailed stellar and nucleosynthetic  models. The models should cover a broad range of stellar masses, initial metallicities, and CNO ratios, 
rotational velocities, and mass loss rates. Regardless the model, it is  necessary that the same basic physics be applied to account 
for the CEMP--no stars, as well as for the other  metal-poor stars,  and for the stars in  the solar neighborhood,  LMC, SMC, and   globular clusters.

$ \quad
\vspace{2mm}
$ 

{\bf{Acknowledgements:}} We thank an anonymous  referee for  very useful remarks that have significantly improved the manuscript.

\bibliographystyle{aa}
\bibliography{Maeder-CEMP}
\end{document}